\documentclass[a4paper,fleqn,usenatbib]{mnras}

\usepackage{txfonts}

\usepackage[T1]{fontenc}
\usepackage{ae,aecompl}
\usepackage{graphics}

\usepackage{psfig}   
\usepackage{graphicx}
\usepackage{amssymb}
\usepackage{subfigure}

\title[H-type objects in IC\,348]{Planet or brown dwarf? Constraints on the formation of H-type objects in IC\,348}

\author[R.~J.~Parker \& C.~Alves~de~Oliveira]{
  Richard J.~Parker\thanks{E-mail: R.Parker@sheffield.ac.uk}\thanks{Royal Society Dorothy Hodgkin Fellow}$^1$ and Catarina Alves de Oliveira$^2$ \vspace*{0.1cm}\\
   $^1$Astrophysics Research Cluster, School of Mathematical and Physical Sciences, The University of Sheffield, Sheffield, S3 7RH, UK \\
   $^2$European Space Agency, European Space Astronomy Centre, Camino Bajo del Castillo s/n, 28692 Villanueva de la Ca{\~n}ada, Madrid, Spain}

\begin{document}

                             
\pagerange{\pageref{firstpage}--\pageref{lastpage}} \pubyear{2026}

\maketitle

\label{firstpage}

\begin{abstract}
The formation mechanism(s) of substellar objects, such as brown dwarfs and free-floating planets, remains an ongoing puzzle in stellar and planetary physics. Recent observational and theoretical work points towards a star-like origin for brown dwarfs, though several authors posit that they could form like planets in a circumstellar disc, and then subsequently be ejected into a star-forming region or the Galactic field. Recently, JWST observations have discovered nine substellar objects in the IC\,348 star-forming region with a spectral absorption feature at 3.4$\mu$m from an unidentified aliphatic hydrocarbon, detected for the first time in planetary atmospheres outside of the Solar System. It is unclear whether  these hydrocarbon absorption features in these `H-type' objects indicate a different formation mechanism compared to more massive brown dwarfs. We quantify the spatial distribution of these objects and find they are indistinguishable from the spatial distribution of stars and other brown dwarfs in IC\,348. We use $N$-body simulations to test whether the H-type objects could have formed as planets in circumstellar discs and then been dynamically ejected by stellar fly-bys. We show that a similar number of free-floating planets could be produced if those planets initially resided at $\sim$\,5\,au from their host stars. However, these free-floating planets have a much more dispersed spatial distribution than the stars and brown dwarfs, inconsistent with the spatial distribution of the H-type objects in IC\,348. We therefore conclude that the H-type objects are unlikely to have a planetary-like origin.     
\end{abstract}

\begin{keywords}   
stars: formation -- brown dwarfs -- kinematics and dynamics -- planets and satellites: gaseous planets -- open clusters and associations: individual (IC\,348) -- methods: numerical
\end{keywords}

\section{Introduction}

The distinction between brown dwarfs and planets has remained an ongoing debate since the discovery of brown dwarfs and exoplanets \citep{Rebolo95,Mayor95}. Do planets exclusively form in circumstellar discs around stars, or are there alternative formation channels, for example the photoerosion of cores \citep{Whitworth04,Gahm07,Haworth15,Diamond24}, or do they form like stars but close to the opacity limit for fragmentation \citep{Bate09,Palau24}?

A major source of confusion is the overlap in mass regimes between brown dwarfs and gas giant planets \citep{Chabrier14,Schlaufman18}. The opacity limit for fragmentation for Solar metallicity gas is around 10\,M$_{\rm Jup}$, making it very difficult to establish whether a several Jupiter-mass object formed like a star, or like a planet. This, and other uncertainties in estimating the masses of young (sub)stellar objects  mean that other diagnostics (such as metallicity/composition or spatial distribution) are sought to distinguish between a stellar-like formation, and a planetary-like formation \citep[e.g.][]{Kumar07,Maldonado17,Parker23c,Giacalone26}.

Recent observations of substellar objects in the IC\,348 star-forming region \citep{Luhman25} with JWST-NIRSpec have shown that some of these objects contain an absorption feature in their spectra at 3.4$\mu$m from an unidentified aliphatic hydrocarbon. This observation constitutes the first detection of this absorption feature in substellar atmospheres outside of the Solar System \citep{Bellucci09}. \citet{Luhman25} proposed a new spectral type for these objects, namely `H-type', to highlight the presence of the hydrocarbon feature in their spectra. The H-type objects have masses between 0.002 -- 0.012\,M$_\odot$, well below the hydrogen burning limit but overlapping with other substellar objects that do not display the 3.4$\mu$m spectral feature.

There are nine H-type objects in IC\,348, and these objects naturally invite a discussion on their origins. Did they form more like stars \citep[through the collapse and fragmentation of the molecular cloud, e.g.][]{Bate09,Stamer19,Palau24}, or did they form as planets in circumstellar discs that have subsequently been ejected through dynamical encounters with passing stars \citep{DaffernPowell22}?

In previous work \citep{Parker23c} we demonstrated that the spatial distribution of planetary-mass objects in NGC\,1333 could be used to contrain the formation mechanism of these objects, as we would expect free-floating planets that formed around stars to have a different spatial (and kinematic) distribution to planetary-mass objects that formed more like stars.

The latest observational census has established that IC\,348 contains 495 stars and brown dwarfs \citep{Luhman16,Luhman24b,Luhman25}, with age determinations for this star-forming region usually between 1 -- 3\,Myr \citep{Muench03,Pavlidou26}, though some authors suggest it may be a factor of two older at around 5 -- 6\,Myr \citep{Bell13,Cottaar14b}. Distance estimates range from 260 -- 270\,pc \citep{Scholz99,Ripepi14} to 320 -- 340\,pc \citep{Cernis93,OrtizLeon18,RuizRodriguez18}, with the most recent determination placing it at 313\,pc \citep{Luhman24b} -- see also \citet{Herbig98} and \citet{Muench07}.

Here, we quantify the spatial distribution of the H-type objects and compare them to the spatial distributions of stars and brown dwarfs in IC\,348.  We then use $N$-body simulations to determine the most likely initial conditions for IC\,348, and the number of free-floating planets we might expect due to dynamical interactions in the region. We then compare the spatial distributions of these free-floating planets to the H-type objects to constrain the formation mechanism of the H-type objects.

The paper is organized as follows. In Section~\ref{sec:observations} we describe the observational dataset. In Section~\ref{sec:methods} we outline our methods. In Section~\ref{sec:results} we present our results and we provide a discussion in Section~\ref{sec:discuss}. We conclude in Section~\ref{sec:conclude}.  

\section{Observational data}
\label{sec:observations}

\begin{figure}
\begin{center}
\rotatebox{270}{\includegraphics[scale=0.35]{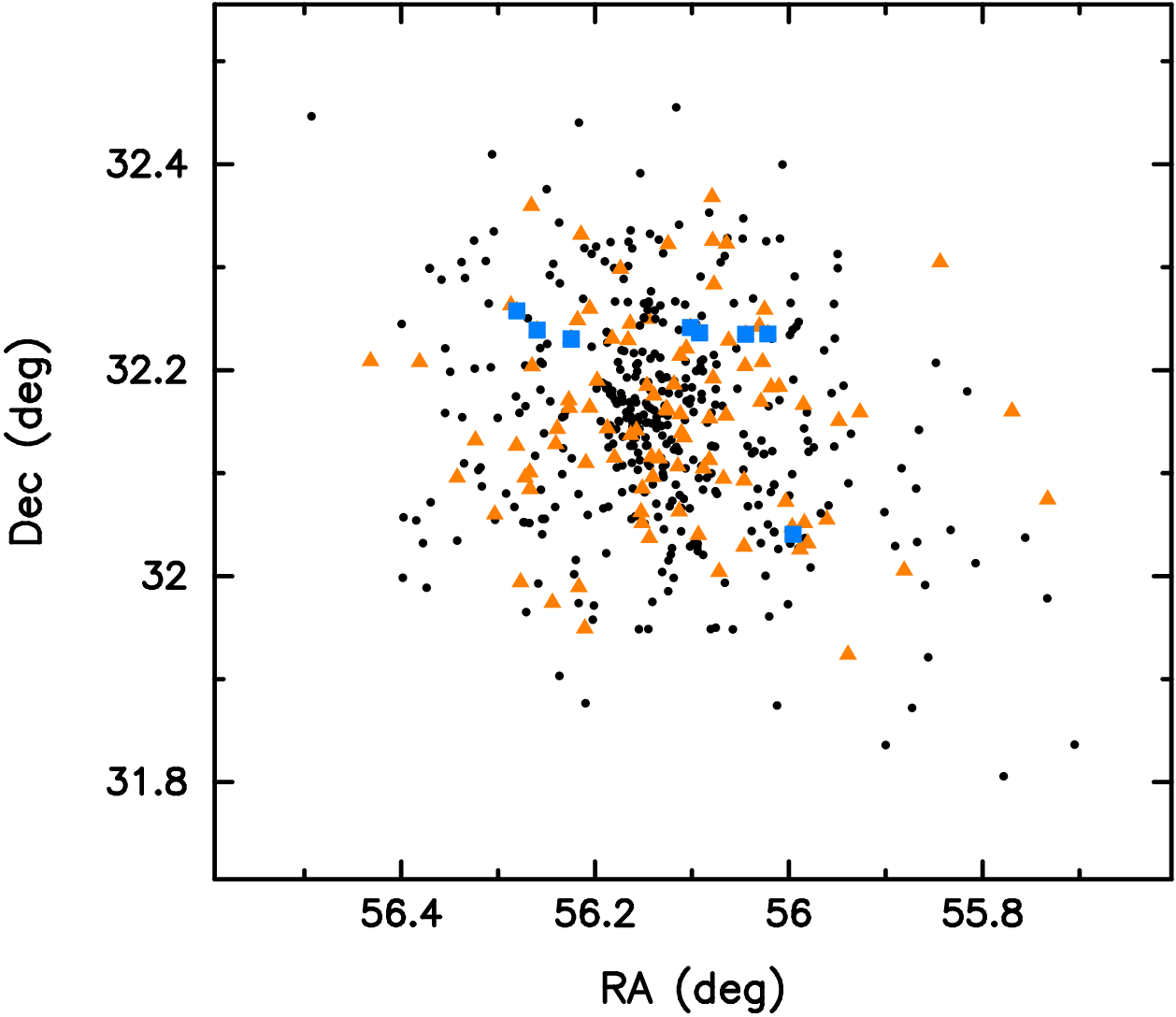}}
\end{center}
\caption[bf]{Map of IC\,348 made using the census of \citet{Luhman16} with updates from \citet{Luhman25}. Positions of stars are shown by the black points. Positions of brown dwarfs are shown by the orange triangles, and the positions of the spectral H-type objects are shown by the blue squares.}
\label{ic348_map}
\end{figure}

Our observational dataset for IC\,348 is the census described in \citet{Luhman16}, supplemented by new additions reported in \citet{Luhman24b} and the new H-type objects reported in \citet{Luhman25}. We plot the positions of these members in Fig.~\ref{ic348_map}, where objects with masses $>$0.1\,M$_\odot$ are shown by the black points, objects with masses less than this are shown by the orange triangles, and the objects with the H-type classification in \citet{Luhman25} are shown by the blue squares.

The object masses are taken from the census described in \citet{Luhman16,Luhman24b} and \citet{Luhman25} and were calculated as follows. Objects with spectral types earlier than K0 have their masses derived from their positions on the Hertzsprung-Russell diagram and evolutionary models as described in \citet{Luhman03b}.  For objects with spectral types between K0 -- M7 their masses were calibrated to dynamical mass estimates, and for objects later than M7 the $M_k$ magnitudes and evolutionary models from \citet{Baraffe15} and \citet{Chabrier23} were used. Based on those models, for pre-main sequence objects in IC\,348, anything later than spectral type M7 is likely to have a mass below the hydrogen-burning limit and is therefore `substellar'. The masses  of the H-type objects  were estimated from a combination of their bolometric luminosities and the evolutionary models.  Luminosities were estimated assuming an adopted distance of 313\,pc \citep{Luhman24b}. 

\section{Methods}
\label{sec:methods}

In this section we first describe the techniques we use to quantify the spatial distributions of the different types of object in both the IC\,348 observational data, and the $N$-body simulation snapshot data. We then describe the set-up and execution of the $N$-body simulations.  

\subsection{Quantifying the spatial distribution of objects}

We use three different measures of the spatial distribution of stars and substellar objects. The first is the local surface density, $\Sigma$, which we calculate using 
\begin{equation}
\Sigma = \frac{N - 1} {\pi r_{N}^2},
\end{equation}
where $r_N$ is the distance to the $N^{\rm th}$ nearest neighbour to the object \citep{Casertano85}. We adopt $N = 10$ (see \citet{Bressert10} and \citet{Parker12d} for a discussion of optimal values to use for $N$). We then plot the local surface density of each object against the object's mass, $m$. The median densities of subsets of objects (categorized by their mass, or some other property such as spectral type) can then be compared to the median density of all objects, and the statistical difference (if any) can be quantified by means of a Kolmogorov-Smirnov (KS) test \citep{Kupper11,Maschberger11,Parker14b}.

In the $N$-body simulations we determine the median surface density of objects $\tilde{\Sigma}$ and plot this at each snapshot of the simulation, such that we can compare the observed value for IC\,348 with the simulations. We use this, and the structure of the star-forming region as quantified by the $\mathcal{Q}$-parameter \citep{Cartwright04,Cartwright09a,Lomax11} to constrain the amount of dynamical evolution that has likely take place within IC\,348 \citep{Parker14b,Parker17a}. We define the $\mathcal{Q}$-parameter in the standard way:
\begin{equation}
\mathcal{Q} = \frac{\bar{m}}{\bar{s}},
\end{equation}
where $\bar{m}$ is the mean length of the Minimum Spanning Tree edge that connects all of the points (stars and substellar objects) in the distribution via a graph with no closed loops. $\bar{s}$ is the mean length of the edge in a complete graph -- where every point in the distribution is connected to every other point. 

We also quantify the spatial distribution of chosen subsets of objects using the $\Lambda_{\rm MSR}$ mass segregation ratio \citep{Allison09a}. This draws a Minimum Spanning Tree between the chosen subset of objects (usually determined by mass) and then compares the length of this MST, $l_{\rm subset}$, to the average length of an MST of randomly drawn objects, $\langle l_{\rm average} \rangle$, which can (randomly) include members of the subset of choice. The ratio of these lengths is the mass segregation ratio, $\Lambda_{\rm MSR}$: 
\begin{equation}
  \Lambda_{\rm MSR} = {\frac{\langle l_{\rm average} \rangle}{l_{\rm subset}}} ^{+ {\sigma_{\rm 5/6}}/{l_{\rm subset}}}_{- {\sigma_{\rm 1/6}}/{l_{\rm subset}}}.
  \label{lambda_msr}
\end{equation}
The uncertainties on the $\Lambda_{\rm MSR}$ ratio are calculated by taking the values $1/6$ and $5/6$ of the way through an ordered list of the MST lengths of the full distribution, and dividing by the length of the subset, such that the lower uncertainty is given by  ${\sigma_{\rm 1/6}}/{l_{\rm subset}}$, and the upper uncertainty is given by ${\sigma_{\rm 5/6}}/{l_{\rm subset}}$ \citep{Parker11b}.

If $\Lambda_{\rm MSR}$ is significantly higher than unity, the subset of objects is mass segregated (closer together than an average sample of objects in the region), whereas if $\Lambda_{\rm MSR}$ is significantly lower than unity, the subset of objects is \emph{inversely} mass segregated (i.e.\,\,more spread out or dispersed than an average sample of objects in the region). As an additional constraint, \citet{Parker15b} and \citet{Parker24} show that whilst values of $0.5 < \Lambda_{\rm MSR} < 2$ are often significant (as determined by the uncertainties), they can still occur through random chance in a small-$N$ (i.e. $N< 1000$) sample, and so we apply the additional constraint when analysing the IC\,348 dataset and our simulations that $\Lambda_{\rm MSR}$ must be both $<0.5$ or $>2$ and significant when the uncertainties are considered.

\subsection{$N$-body simulations}

There are 495 stellar and substellar objects in the \citet{Luhman16,Luhman25} sample, and we draw $N = 495$ objects from a \citet{Maschberger13} Initial Mass Function of the form
\begin{equation}
p(m) \propto \left(\frac{m}{\mu}\right)^{-\alpha}\left(1 + \left(\frac{m}{\mu}\right)^{1 - \alpha}\right)^{-\beta}.
\label{maschberger_imf}
\end{equation}
In Eqn.~\ref{maschberger_imf}  $\mu = 0.2$\,M$_\odot$ is the scale parameter, or `peak' of the IMF \citep{Bastian10,Maschberger13}, $\alpha = 2.3$ is the \citet{Salpeter55} power-law exponent for higher mass stars, and $\beta = 1.4$ describes the slope of the IMF for low-mass objects. We randomly sample this distribution in the mass range 0.001 -- 50\,M$_\odot$, such that we sample objects down to the planetary mass regime. Objects with masses between 0.001 and 0.08\,M$_\odot$ are brown dwarfs, and are assumed to form in the same way as stars. 

In addition to the objects drawn from the IMF (which we consider to have formed from the collapse and fragmentation of the giant molecular cloud), we also place a 1\,M$_{\rm Jup}$ planet in orbit around every star in the mass range 0.1 -- 3\,M$_\odot$. We assume every star is able to form a Jupiter-mass planet, which may not be representative of all planetary systems. This choice is simply to increase the potential number of free-floating massive planets.  We also assume that the planets form instantaneously around the stars, which is likely to be unrealistic, though observations do suggest Jupiter-mass planets can form within short timescales of the order 0.5 -- 1\,Myr \citep{Alves20,SeguraCox20}.

For each star-forming region we run the simulation three times where we change the initial semimajor axis of the planets; in one simulation they are all placed at 1\,au, in another they are at 5\,au and in the final simulation they are at 30\,au. The initial eccentricities of the planets are set to zero. The semimajor axes are chosen simply because they represent the positions of Earth, Jupiter and Neptune in our Solar System; as we will discuss in Section~\ref{sec:results}, using discretized values simplifies the interpretation of the numbers of free-floating planets that can be produced through dynamical encounters as a function of the stellar density, without choosing a fully populated planet semimajor axis distribution.

These objects, which all have mass 1\,M$_{\rm Jup} = 9.4 \times 10^{-4}$\,M$_\odot$ are assumed to have formed like planets in circumstellar discs. If they are liberated from their parent star due to dynamical interactions, we label them `free-floating planets', to distinguish them from the slightly more massive brown dwarfs. 

Observations \citep[e.g.][]{Larson81,Gomez93,Larson95,Cartwright04,Sanchez09,Kuhn14,Henshaw16b,Hacar18,Buckner19,Kuhn19} and simulations \citep[e.g.][]{Schmeja06,Girichidis11} show that star-forming regions initially form with a high degree of spatial and kinematic substructure, which is then erased due to dynamical encounters and relaxation \citep[e.g.][]{Goodwin04a,Parker14b}. 

We use the box-fractal method \citep{Goodwin04a,Lomax18,DaffernPowell20} to set up substructured star-forming regions. For a full description of the method we refer the reader to \citet{DaffernPowell20} but we briefly restate it here.

We start by defining a cube whose sides have a length $N_{\rm div} = 2$. We place a `root' particle at the centre of the cube, and the cube is then divided into  $N^3_{\rm div}$ sub-cubes, and a `leaf' particle is placed at the centre of each of these sub-cubes.

The probability that each leaf particle will become a root particle is given by $N^{(D - 3)}_{\rm div}$, where $D$ is the fractal dimension. In three dimensions, when $D = 1.6$ few leaf particles become root particles, and the distribution has a highly substructured appearance. When $D = 3.0$ all of the leaf particles themselves become root particles and so the distribution has a smooth appearance.

The final leaf particles have a small amount of noise added to their positions, to prevent the distribution having a gridded appearance. The procedure sometimes produces more particles than are required, and the excess particles are randomly pruned whilst preserving the desired fractal dimension $D$. 

The velocity structure is also determined by the root+leaf particle generation above. The first root star has a velocity drawn from a Gaussian with mean of zero, and then every leaf star inherits this velocity plus an additional random component drawn from the same Gaussian but multiplied by $\left(1/N_{\rm div}\right)^g$, where $g$ is the number of the root+leaf generation the particle was produced in.

This way of generating velocities means that physically close particles have similar velocities, but those that are distant from one another may have very different velocities. This is physically motivated by the \citet{Larson81} relations, where nearby cores/protostars have small velocity dispersions, and distant cores/proto stars have larger velocity dispersions. 

We then scale the velocities to the virial ratio of the star-forming region, $\alpha_{\rm vir} = T/|\Omega|$, where $T$ and $|\Omega|$ are the total kinetic and total potential energies of all the objects, respectively. Again, to replicate observations of the virial ratio of cores and proto-stars in star-forming regions \citep[e.g.][]{Foster15}, we set the virial ratio to be subvirial, with $\alpha_{\rm vir} = 0.3$.

Different combinations of the fractal dimension, $D$ and the radius $r_F$ of each region, are used to produce regions with a range of different initial stellar densities. We summarise all of the combinations (and resultant stellar densities) in Table~\ref{initial_conditions}.

To obtain an idea of the stochasticity in the simulations, we run ten versions of each set of initial conditions, identical apart from the random number seed used to initialise the positions, masses and velocities of the objects. We evolve each simulation for 10\,Myr using the $4^{\rm th}$-order Hermite $N$-body integrator \texttt{kira} within the \texttt{Starlab}  environment \citep{Zwart99,Zwart01}. We do not include stellar evolution in, or impose an external tidal field on, the simulations.

\begin{table*}
\caption[bf]{A summary of the initial conditions in our simulations. We show the initial radius $r_F$, fractal dimension $D$, the range of initial stellar densities $\tilde{\rho}_{\rm 0\,Myr}$ these radii and fractal dimensions lead to, the corresponding surface density range $\tilde{\Sigma}_{\rm 0\,Myr}$ and the morphology $\mathcal{Q}_{\rm 0\,Myr}$ range. We also provide the range of stellar surface densities and morphologies at 1, 3 and 5\,Myr following dynamical evolution. Simulations where the surface density and structure are both consistent with the observed values for IC\,348 are underlined.}
\begin{center}
  \begin{tabular}{|c|c|c|c|c|c|c|c|c|c|c|}
    \hline
    $r_F$ & $D$ & $\tilde{\rho}_{\rm 0\,Myr}$ & $\tilde{\Sigma}_{\rm 0\,Myr}$ & $\mathcal{Q}_{\rm 0\,Myr}$ & $\tilde{\Sigma}_{\rm 1\,Myr}$ & $\mathcal{Q}_{\rm 1\,Myr}$ & $\tilde{\Sigma}_{\rm 3\,Myr}$ & $\mathcal{Q}_{\rm 3\,Myr}$ & $\tilde{\Sigma}_{\rm 5\,Myr}$ & $\mathcal{Q}_{\rm 5\,Myr}$ \\
    (pc) & & (M$_\odot$\,pc$^{-3}$) & (stars\,pc$^{-2}$) &  & (stars\,pc$^{-2}$) & & (stars\,pc$^{-2}$) & & (stars\,pc$^{-2}$) & \\

    \hline
    0.5 & 1.6 & $0.7 - 5 \times 10^5$ & $0.5 - 2 \times 10^4$ & $0.25 - 0.35$ & $0.5 - 3 \times 10^3$ & $1.25 - 1.75$ & $0.8 - 2 \times 10^2$ & $1.10 - 1.25$ &$0.2 - 1 \times 10^2$ &$1.05 - 1.20$ \\
    1.0 & 1.6 & $0.9 - 6 \times 10^4$ & $1 - 4 \times 10^3$ & $0.25 - 0.35$ & $0.6 - 2 \times 10^3$ & $1.05 - 1.45$ & $1 - 4 \times 10^2$ & $1.05 - 1.35$ & $0.4 - 2 \times 10^2$ & $1.00 - 1.15$\\ 
    1.3 & 1.6 & $0.4 - 3 \times 10^4$ & $0.8 - 3 \times 10^3$& $0.25 - 0.35$ & $0.5 - 2 \times 10^3$& $0.85 - 1.18$ & \underline{\bf $0.9 - 4 \times 10^2$} & \underline{\bf $0.95 - 1.28$} & \underline{\bf $0.4 - 2 \times 10^2$} & \underline{\bf $0.90 - 1.16$} \\ 
    1.5 & 1.6 & $0.3 - 2 \times 10^4$ & $0.6 - 2 \times 10^3$ &$0.25 - 0.35$ & $0.5 - 1 \times 10^3$ & $0.75 - 1.08$ & $2 - 6 \times 10^2$ & $0.95 - 1.30$ & $0.6 - 3 \times 10^2$ & $1.05 - 1.22$ \\ 
    2.0 & 1.6 & $1 - 8 \times 10^3$ & $0.3 - 1 \times 10^3$ & $0.25 - 0.35$ & $0.2 - 1 \times 10^3$ & $0.40 - 0.80$& $2 - 5 \times 10^2$ & $0.70 - 1.10$ & \underline{\bf $0.8 - 2 \times 10^2$} & \underline{\bf $0.95 - 1.17$} \\
    \hline
    0.5 & 2.0 & $1.5 - 2.5 \times 10^4$ & $3 - 4 \times 10^3$ & $0.35 - 0.50$ & $0.8 - 3 \times 10^3$ & $1.20 - 1.65$ & $1 - 4 \times 10^2$ & $1.15 - 1.30$ & $0.5 - 2 \times 10^2$ & $1.12 - 1.23$ \\
    1.0 & 2.0 & $2 - 4 \times 10^3$ & $0.6 - 1 \times 10^3$ & $0.35 - 0.50$ & $0.5 - 1 \times 10^3$ & $0.85 - 1.20$ & $ 1 - 7 \times 10^2$ & $1.05 - 1.30$ & $0.6 - 3 \times 10^2$ & $1.05 - 1.20$ \\ 
    1.3 & 2.0 & $0.8 - 1.6 \times 10^3$ & $4 - 7 \times 10^2$ & $0.35 - 0.50$ & $3 - 7 \times 10^2$ & $0.68 - 1.00$ & $2 - 9 \times 10^2$ & $0.95 - 1.18$ & $0.7 - 2 \times 10^2$ & $1.05 - 1.16$ \\ 
    1.5 & 2.0 & $0.5 - 1 \times 10^3$ & $3 - 5 \times 10^2$ & $0.35 - 0.50$  & $2 - 5 \times 10^2$ & $0.58 - 0.81$ & $2 - 8 \times 10^2$ & $0.90 - 1.20$ & $0.7 - 3 \times 10^2$ & $0.99 - 1.17$ \\ 
    2.0 & 2.0 & $2.5 - 4 \times 10^2$ & $2 - 3 \times 10^2$ & $0.35 - 0.50$ & $1 - 3 \times 10^2$  & $0.52 - 0.72$ & $1.5 - 4 \times 10^2$ & $0.75 - 0.95$ & $1 - 4 \times 10^2$ & $0.93 - 1.17$\\
    \hline
    0.5 & 3.0 & $2 - 3 \times 10^3$ & $1 - 2 \times 10^3$ &$0.70 - 0.80$ & $0.8 - 3 \times 10^3$ & $0.97 - 1.20$ & $1 - 8 \times 10^2$ & $1.08 - 1.40$ & $0.4 - 3 \times 10^2$ & $1.09 - 1.28$\\
    2.0 & 3.0 & $0.8 - 1 \times 10^2$ & $1 - 2 \times 10^2$ &$0.70 - 0.80$ &$1.5 - 2.5 \times 10^2$ & $0.80 - 0.90$ & $3 - 5 \times 10^2$ & $0.89 - 1.23$ & \underline{\bf $0.7 - 3 \times 10^2$} & \underline{\bf $0.96 - 1.08$}\\
    \hline
  \end{tabular}
\end{center}
\label{initial_conditions}
\end{table*}

\section{Results}
\label{sec:results}

In this Section we first quantify the spatial distribution of the H-type objects in IC\,348, before constraining the likely initial conditions for IC\,348 using $N$-body simulations. We then use these $N$-body simulations to calculate the number of free-floating planets that would be produced solely through stellar fly-bys in these star-forming regions. We then quantify the spatial distribution of the free-floating planets that remain within the star-forming regions after they are ejected. 

\subsection{The spatial distribution of H-type objects in IC\,348}

First, we calculate the $\mathcal{Q}$-parameter on the updated dateset from \citet{Luhman25} and find exactly the same values ($\bar{m} = 0.355$, $\bar{s} = 0.360$, $\mathcal{Q} = 0.98$) as we reported in \citet{Parker17a}.

\begin{figure}
\begin{center}
\rotatebox{270}{\includegraphics[scale=0.35]{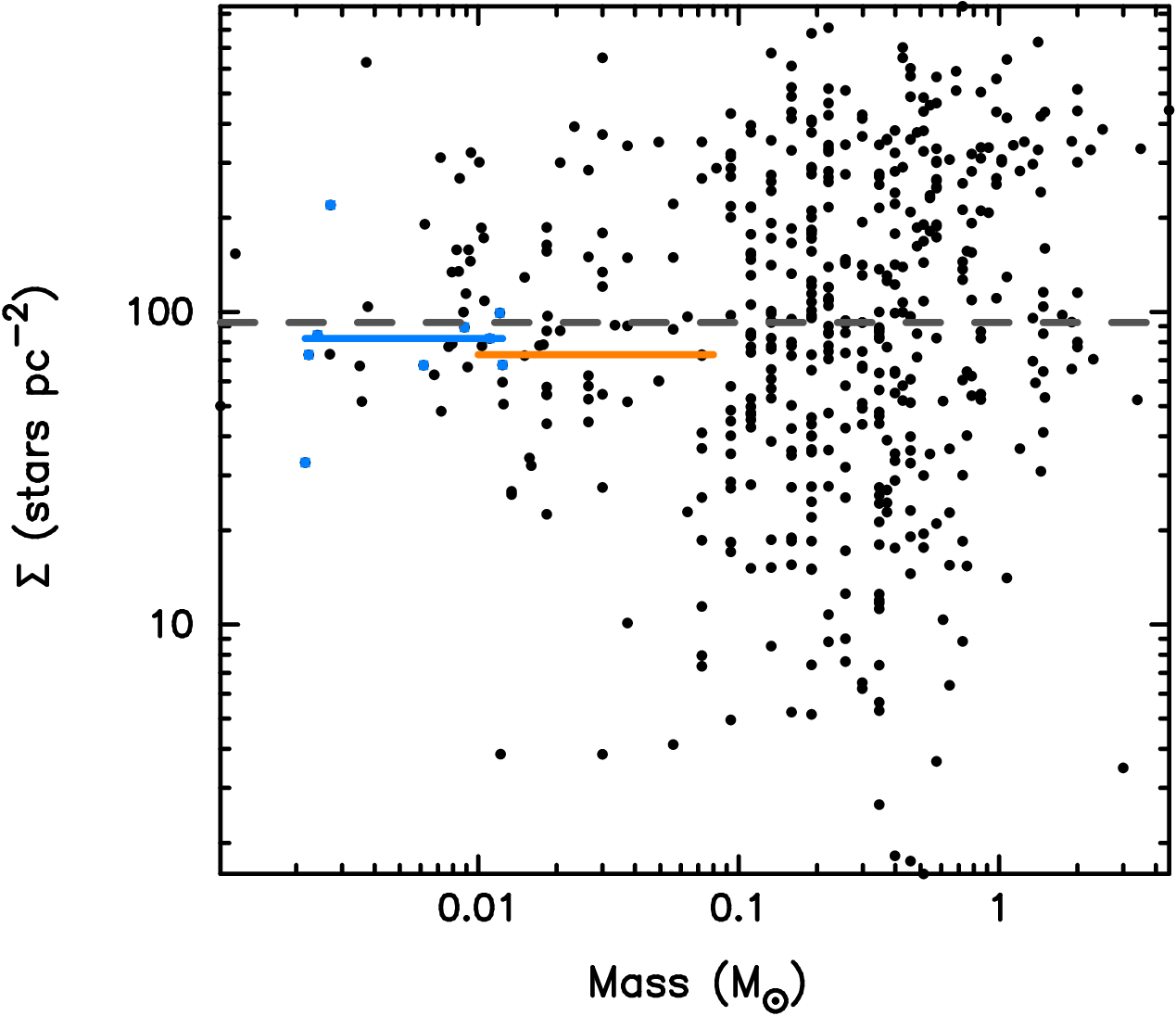}}
\end{center}
\caption[bf]{Local surface density calculated using the ten nearest neighbours to each object plotted against the object's mass for the IC\,348 star-forming region. The spectral H-type objects are shown by the blue squares, and the median surface density of the nine H-type objects is shown by the blue horizontal line (which also shows the mass range of these objects). The median local surface density of brown dwarfs in the mass range 0.01 -- 0.08\,M$_\odot$ is shown by the orange horizontal line. The median local density for all objects is shown by the grey dashed line.  }
\label{ic348_Sigma-m}
\end{figure}

We show the local surface density of all of the objects in IC\,348 and plot them against their mass in Fig.~\ref{ic348_Sigma-m}. The median surface density of all objects is $95^{+135}_{-79}$\,stars\,pc$^{-2}$, assuming a distance of 313\,pc \citep{Luhman24b}, and the uncertainty on this surface density comes from the lowest (261\,pc) and highest (340\,pc) estimates of the distance to IC\,348 \citep{Cernis93,Scholz99}. The H-type objects are plotted with blue squares, and we indicate their median surface density by the horizontal blue line, which is 83\,stars\,pc$^{-2}$. In comparison, the median local density of brown dwarfs in the mass range 0.01 -- 0.08\,M$_\odot$ is 74\,stars\,pc$^{-2}$ and is shown by the horizontal orange line, and the median local surface density of all objects (including brown dwarfs and the H-types) is shown by the horizontal dashed line.

We perform KS tests between the full sample of objects and the H-type objects ($p$-value 0.16), and the full sample of objects and the brown dwarfs ($p$-value 0.12). In both tests we therefore cannot reject the hypothesis that the two samples contain the same underlying parent density distribution.

For the $\Sigma - m$ plot in Fig.~\ref{ic348_Sigma-m}, we are able to accomodate the fact that the H-type objects have overlapping masses with the brown dwarfs. To determine whether the H-type objects have a different spatial distribution from the other objects in IC\,348 with $\Lambda_{\rm MSR}$, we must distinguish them from the brown dwarfs with similar masses.

\begin{figure*}
  \begin{center}
\setlength{\subfigcapskip}{10pt}
\hspace*{-1.5cm} \subfigure[]{\label{IC348_Lambda-a}\rotatebox{270}{\includegraphics[scale=0.35]{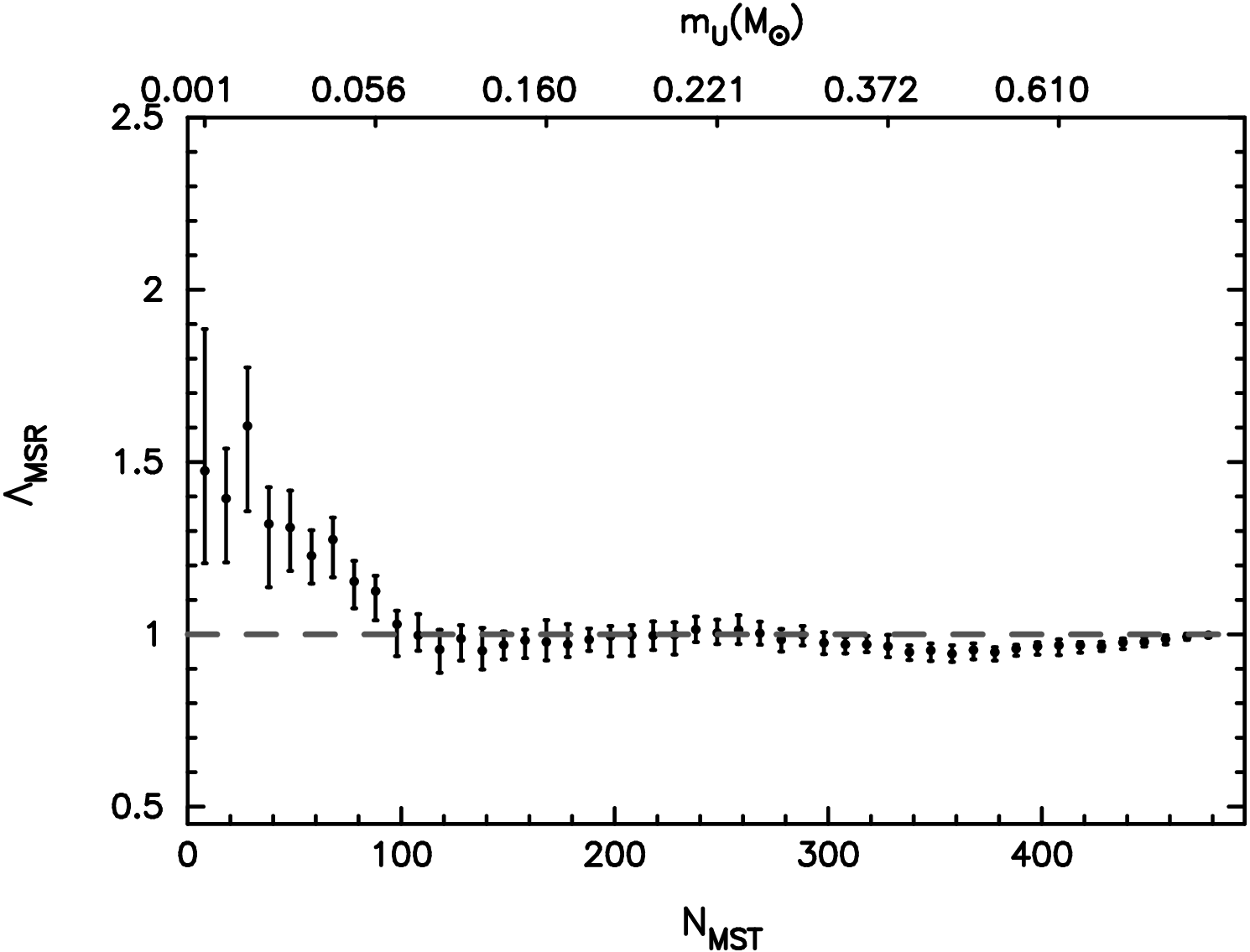}}} 
\hspace*{0.3cm}\subfigure[]{\label{IC348_Lambda-b}\rotatebox{270}{\includegraphics[scale=0.35]{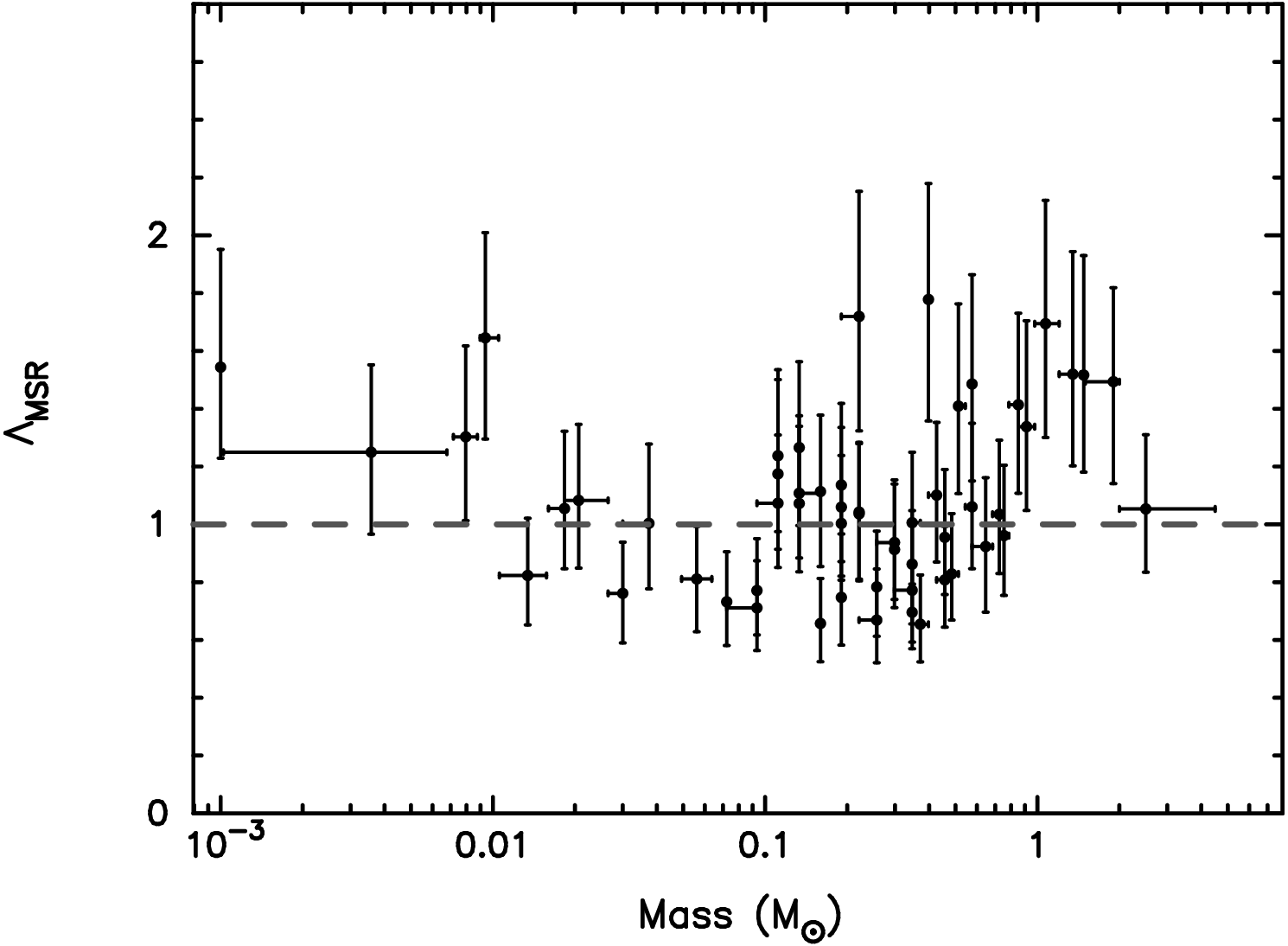}}}

\caption[bf]{The mass segregation ratio, $\Lambda_{\rm MSR}$, for the lowest-mass objects in IC\,348, where we have assigned all of the H-type objects a mass of $1 \times 10^{-3}$\,M$_\odot$. In panel (a) we show the $\Lambda_{\rm MSR}$ ratio as a function of the $N_{\rm MST}$ least massive objects -- first constructing a bin containing the nine least-massive objects (all of the H-type objects) and then adding the next nine next-least massive objects. In panel (b) we show the   $\Lambda_{\rm MSR}$ ratio for fixed bins containing nine objects. Here, the error bars in the $x$-axis are the mass range of each bin. In both panels, the error bars in the $y$-axis are the uncertainties in the determination of $\Lambda_{\rm MSR}$, as described in the text. The horizontal dashed line indicates $\Lambda_{\rm MSR} = 1$, i.e. no mass segregation.    }
\label{IC348_Lambda}
  \end{center}
\end{figure*}

In Fig.~\ref{IC348_Lambda} we show the $\Lambda_{\rm MSR}$ mass segregation ratio, where we have set all of the H-type masses to be $1 \times 10^{-3}$\,M$_\odot$. This enables us to easily group the H-type objects together in the same bin(s) in order to quantify whether their mass segregation ratio significantly deviates from unity (i.e. not mass segregated).

We show the $\Lambda_{\rm MSR}$ mass segregation ratio for the lowest mass objects (the H-types) in Fig.~\ref{IC348_Lambda-a}. The leftmost datapoint contains just the nine H-type objects, and then subsequent bins contain the next ten least massive objects, and so on. The $\Lambda_{\rm MSR}$ mass segregation ratio for the H-type objects is slightly above unity, though \citet{Parker15a} show that values less than two are unlikely to be significant, as they can simply occur from randomly placing a low number of stars in a spatial configuration.

In Fig.~\ref{IC348_Lambda-b} we show the $\Lambda_{\rm MSR}$ mass segregation ratio for independent bins of nine objects. Again, the H-type objects (leftmost bin) have a mass segregation ratio slightly above unity, but not at a significantly higher level than any of the more massive objects in the star-forming region.

To check that artificially assigning masses of $1 \times 10^{-3}$M\,$_\odot$ to the H-type objects does not affect the determination of $\Lambda_{\rm MSR}$ we show a similar figure in the Appendix (Fig.~\ref{IC348_Lambda_all_masses}) where the H-type objects retain the masses assigned to them by \citet{Luhman25}.\\

In summary, the H-type objects in IC\,348 do not have a different spatial distribution to the other members of the region, either in terms of their local surface densities or the typical distances between other H-type objects, and randomly selected objects in the star-forming region. 

\subsection{Free-floating planets in $N$-body simulations}

We now test the hypothesis that the H-type objects in IC\,348 are planetary objects that formed around stars and were subsequently liberated from their parent stars by a flyby with a passing star. The number of free-floating planets produced in this way is strongly dependent on the initial conditions of the star-forming region.

We first establish the most likely initial conditions for IC\,348 by matching the spatial structure (quantified by the $\mathcal{Q}$-parameter), and the average stellar surface density, of IC\,348 to $N$-body simulations. Each snapshot from a simulation where the initial conditions set produces values of both $\tilde{\Sigma}$ and $\mathcal{Q}$ that are consistent with the observations is underlined in Table~\ref{initial_conditions}.

We determine that a simulation is consistent with the observed values if the observed surface density 95\,stars\,pc$^{-2}$ and $\mathcal{Q}$-parameter ($\mathcal{Q}$ = 0.98) is within the range of simulation values  after 1 -- 5\,Myr of dynamical evolution in the simulation. This age range accounts for the uncertainties surrounding the age of IC\,348. The stellar density condition could also be relaxed if we allow the simulations to be within the observed uncertainties on the surface density (which in turn come from the uncertainty on the distance determination).

\begin{figure*}
  \begin{center}
\setlength{\subfigcapskip}{10pt}
\hspace*{-1.5cm} \subfigure[]{\label{density_structure-a}\rotatebox{270}{\includegraphics[scale=0.35]{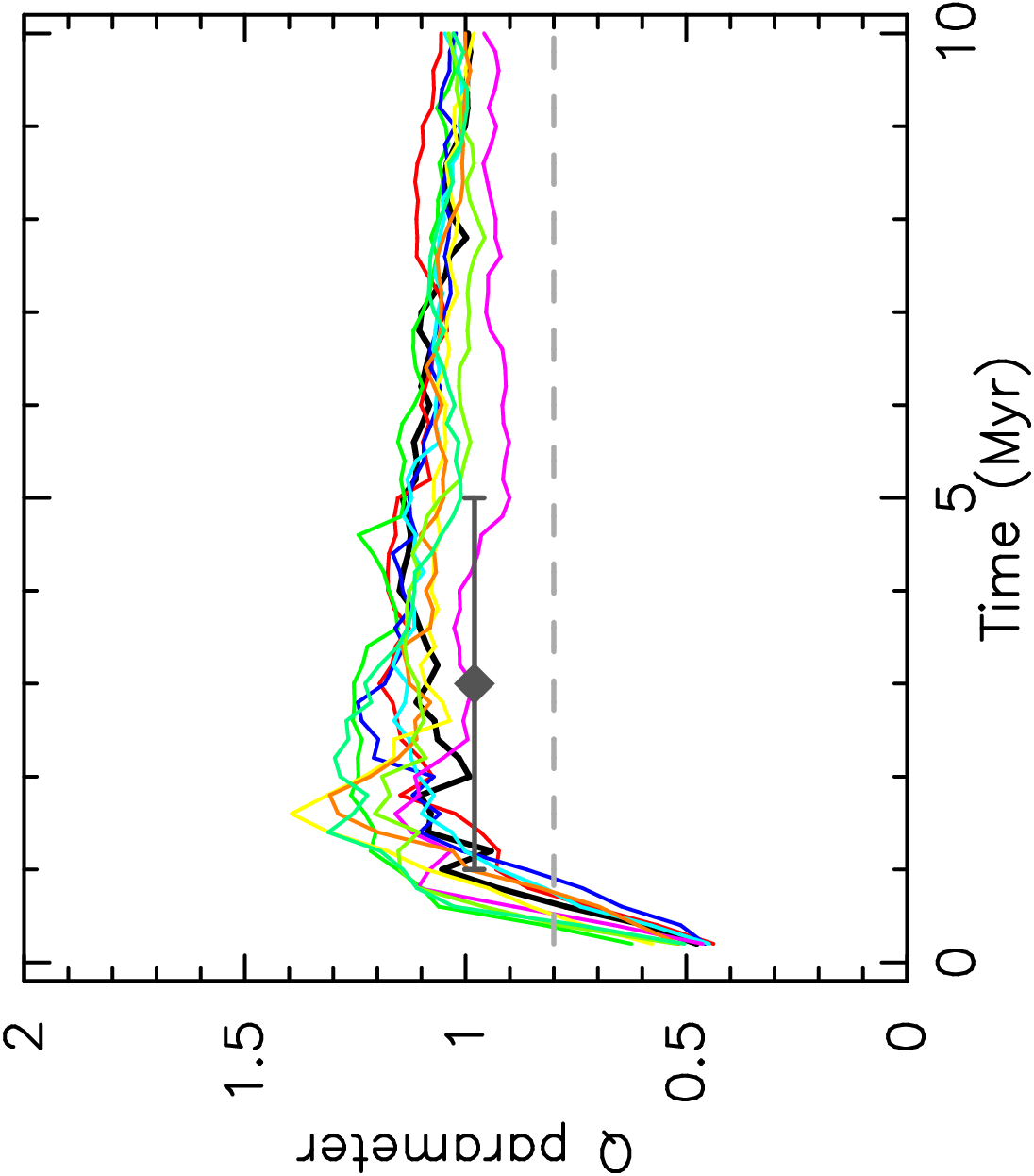}}} 
\hspace*{0.5cm}\subfigure[]{\label{density_structure-b}\rotatebox{270}{\includegraphics[scale=0.35]{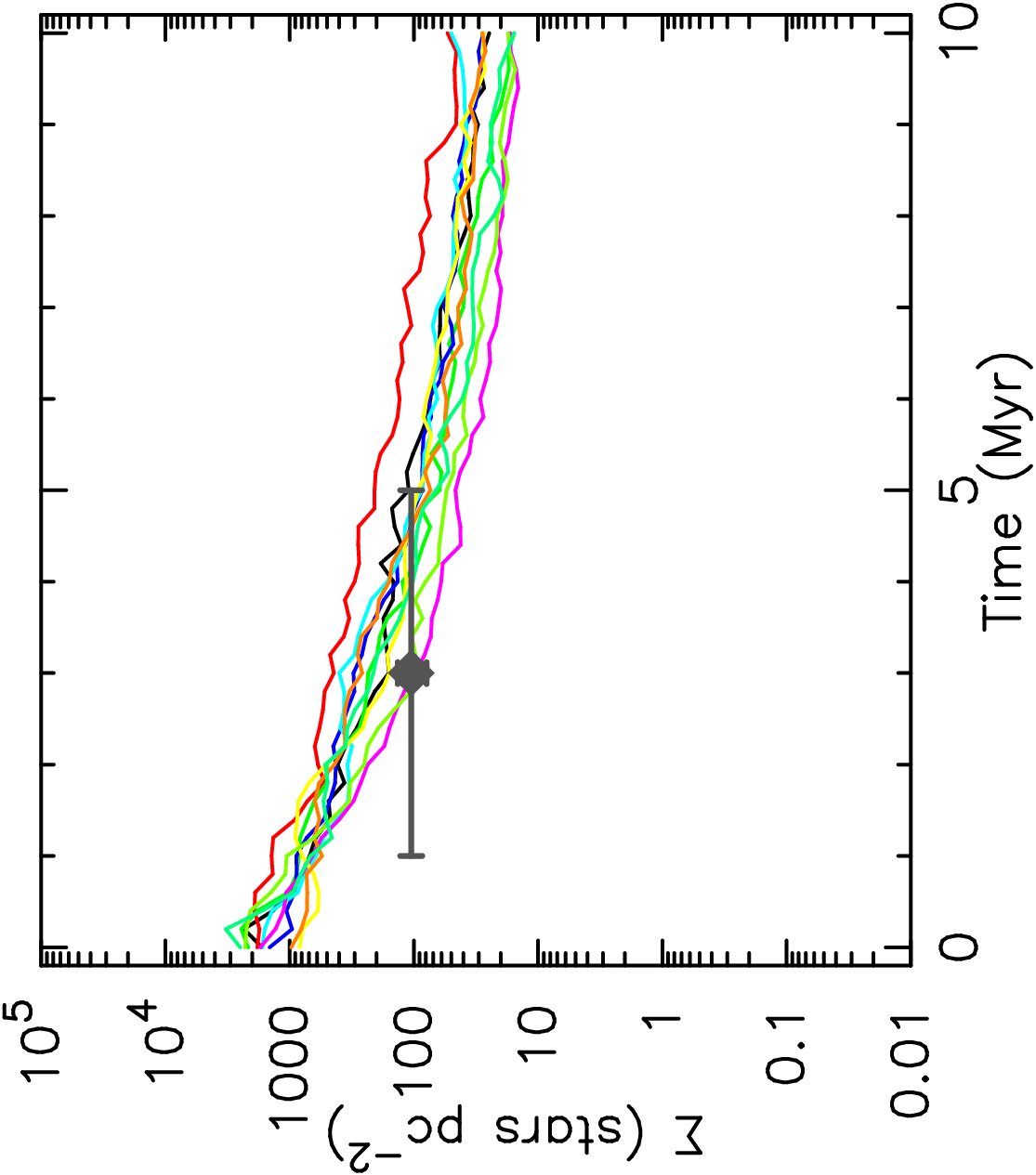}}}

\caption[bf]{$N$-body simulations showing the evolution of morphology and stellar density over time. Each coloured line represents a different realisation of the same initial conditions for our best fitting simulation set, which has an initial radius of $r_F = 1.3$\,pc and a fractal dimension $D = 1.6$. Panel (a) shows the evolution of structure as quantified by the $\mathcal{Q}$-parameter, and panel (b) shows the evolution of the median stellar surface density, $\tilde{\Sigma}$. The observed values for IC\,348 (with large error bars to highlight the uncertainty on the age of this star forming-region) are shown by the diamond symbols. In panel (a) the horizontal dashed line shows the boundary between substructured ($\mathcal{Q} < 0.8$) and smooth ($\mathcal{Q} > 0.8$) morphologies.}
\label{density_structure}
  \end{center}
\end{figure*}

The full evolution of our best-fitting simulation set is shown in Fig.~\ref{density_structure}. We take a `good' fit as one where both the $\mathcal{Q}$-parameter (Fig.~\ref{density_structure-a}) and the median local surface density (Fig.~\ref{density_structure-b}) are consistent with the observations. In both panels the diamond symbol is the observed value for IC\,348, assuming an age of 3\,Myr (with the horizontal error bars indicating the uncertainty in the age of the region).

The best fitting simulation set has an initial radius of $r_F = 1.3$\,pc and a fractal dimension $D = 1.6$. If we increase the initial density (decrease the radius), the $\mathcal{Q}$-parameter is too high after 3\,Myr, and if we reduce the initial density the $\mathcal{Q}$-parameter is too low.

Two other simulation sets ($r_F = 2.0$, $D = 1.6$ and $r_F = 2.0$, $D = 3.0$) are consistent with the observations of IC\,348, but only if the age of IC\,348 is closer to 5\,Myr.

We then use the best-fitting initial conditions to calculate the number of free-floating planets we expect in these star-forming regions. We run three versions of the same simulation, placing the planets at different semimajor axes (1, 5 and 30\,au) in each simulation.

\begin{figure*}
  \begin{center}
\setlength{\subfigcapskip}{10pt}
\hspace*{-1.5cm} \subfigure[$a_p = 1$\,au]{\label{FFLOP_numbers-a}\rotatebox{270}{\includegraphics[scale=0.28]{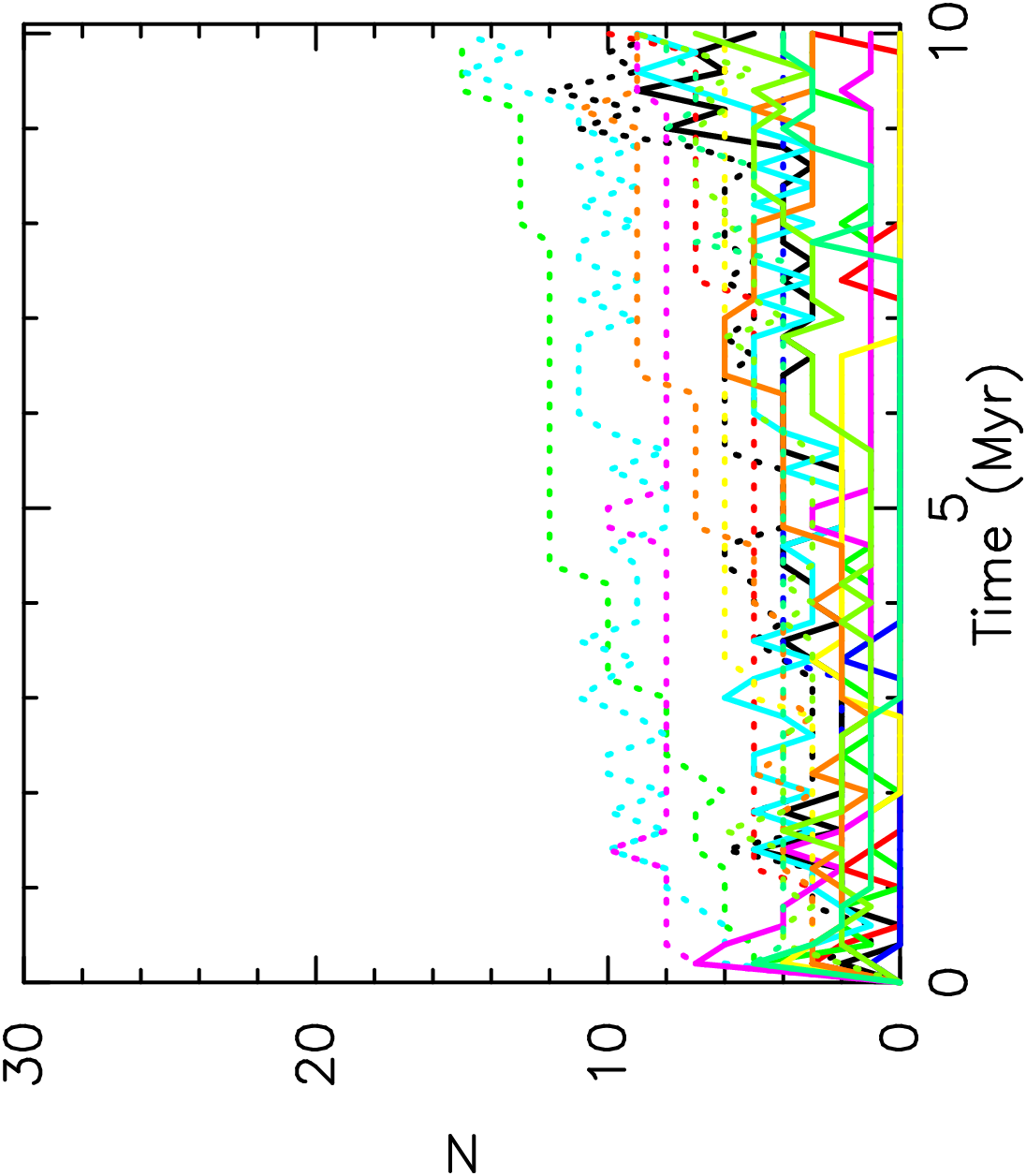}}} 
\hspace*{0.3cm}\subfigure[$a_p = 5$\,au]{\label{FFLOP_numbers-b}\rotatebox{270}{\includegraphics[scale=0.28]{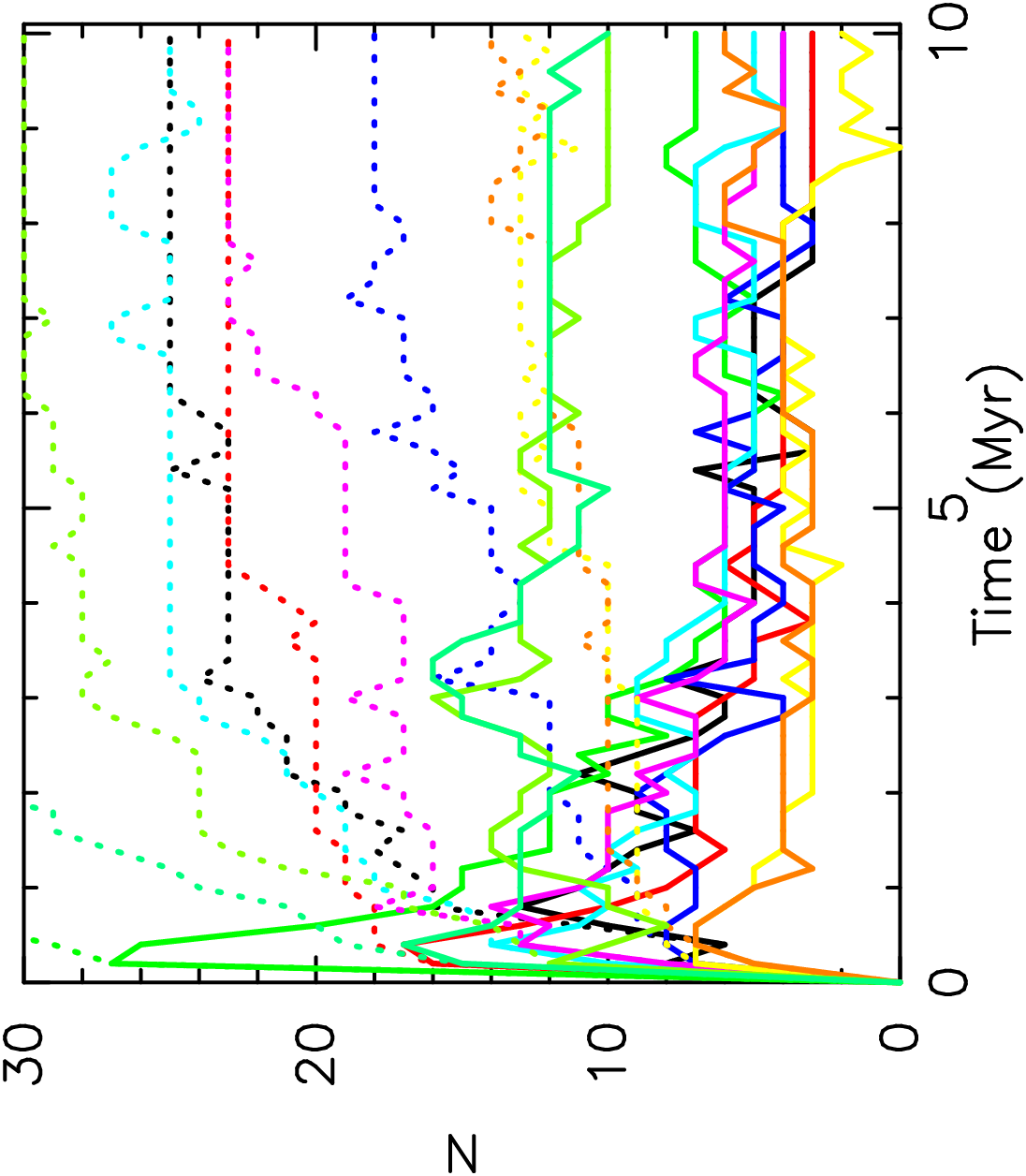}}}
\hspace*{0.3cm}\subfigure[$a_p = 30$\,au]{\label{FFLOP_numbers-c}\rotatebox{270}{\includegraphics[scale=0.28]{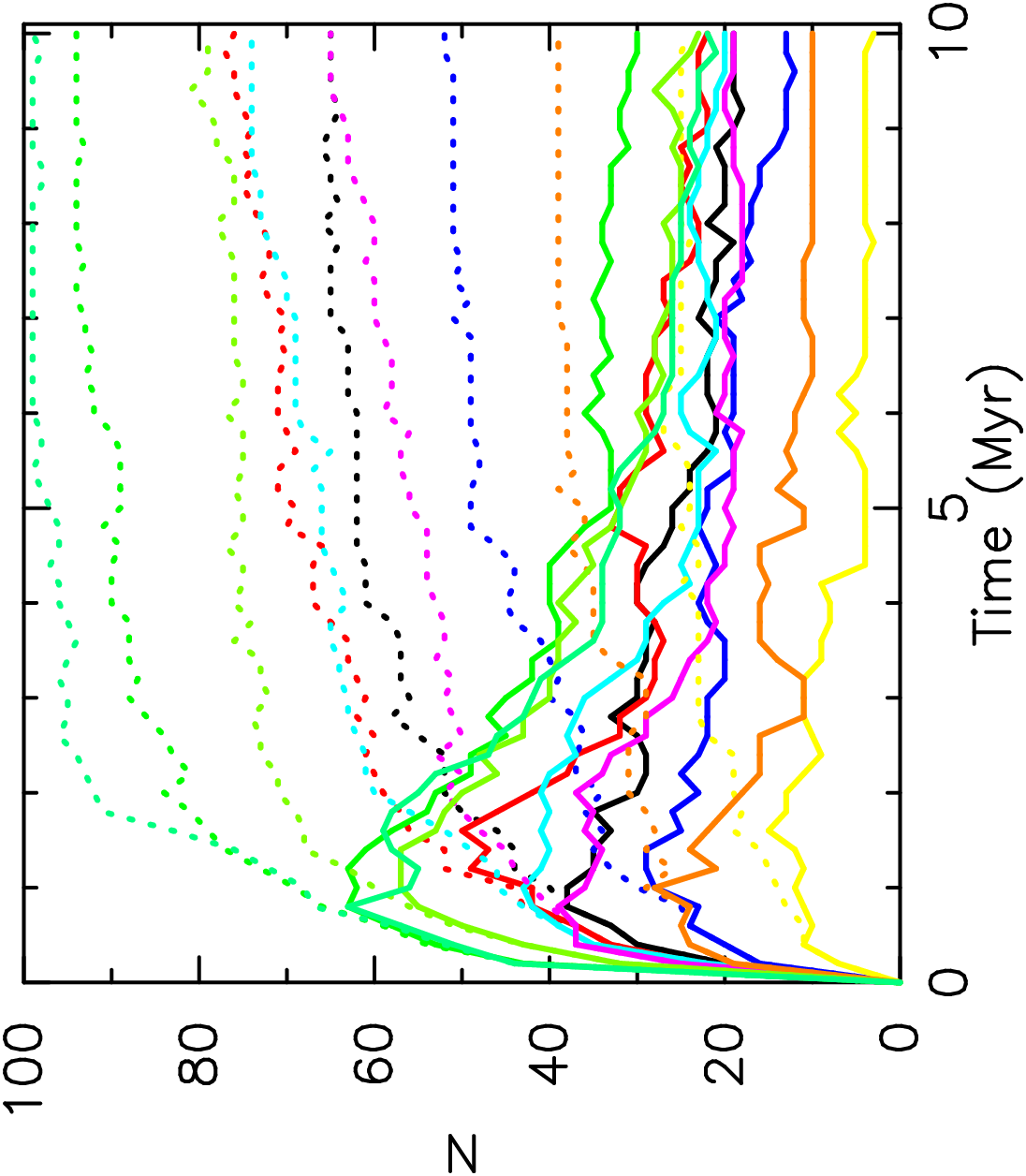}}}

\caption[bf]{The evolution of the number of free-floating planets in our simulations that best match the present-day morphology and median surface density of IC\,348. Each coloured line shows the number of free-floating planets -- those liberated from their host stars -- in simulations where the initial conditions are statistically similar. The solid lines show the number of planets that remain within 5\,pc of the centre of the star-forming region, and the dotted lines show the total number of free-floating planets. The panels show versions of the same simulations set but where the planets are all initially at 1\,au (panel (a)), 5\,au (panel(b)) and 30\,au (panel (c)).  Note the different $y$-axis scale for the simulations with planets initially at 30\,au (panel (c)), which is due to these planets having a lower binding energy and hence being easier to liberate in an interaction. }
\label{FFLOP_numbers}
  \end{center}
\end{figure*}

We show the number of free-floating planets produced in each star-forming region in Fig.~\ref{FFLOP_numbers}. The dotted lines show the \emph{total} numbers of free-floating planets produced in each simulation, and the solid lines show the total number of free-floating planets that remain within 5\,pc of the centre of the star-forming region. We adopt 5\,pc as objects beyond this distance from the centre of IC\,348 may not be included in the observational sample, even if they originated in the region and were ejected. Each coloured line represents a different realisation of the same initial conditions.

If we assume the H-type objects are free-floating planets, and there are nine in IC\,348, then we select the simulations that produce between 5 and 15 that remain within 5\,pc of the centre of the star-forming region. Only one simulation in which the planets are originally at 1\,au (Fig.~\ref{FFLOP_numbers-a}) produces this amount of free-floating planets, and 8 out of 10 simulations where the planets are originally at 30\,au produce too many free-floating planets (Fig.~\ref{FFLOP_numbers-c}). The simulations where the planets are all originally at 5\,au (Fig.~\ref{FFLOP_numbers-b}) produce similar numbers of planets to the number of H-type objects in eight out of ten simulations.

\begin{figure*}
  \begin{center}
\setlength{\subfigcapskip}{10pt}
\hspace*{-1.5cm} \subfigure[]{\label{Sim_05-a}\rotatebox{270}{\includegraphics[scale=0.35]{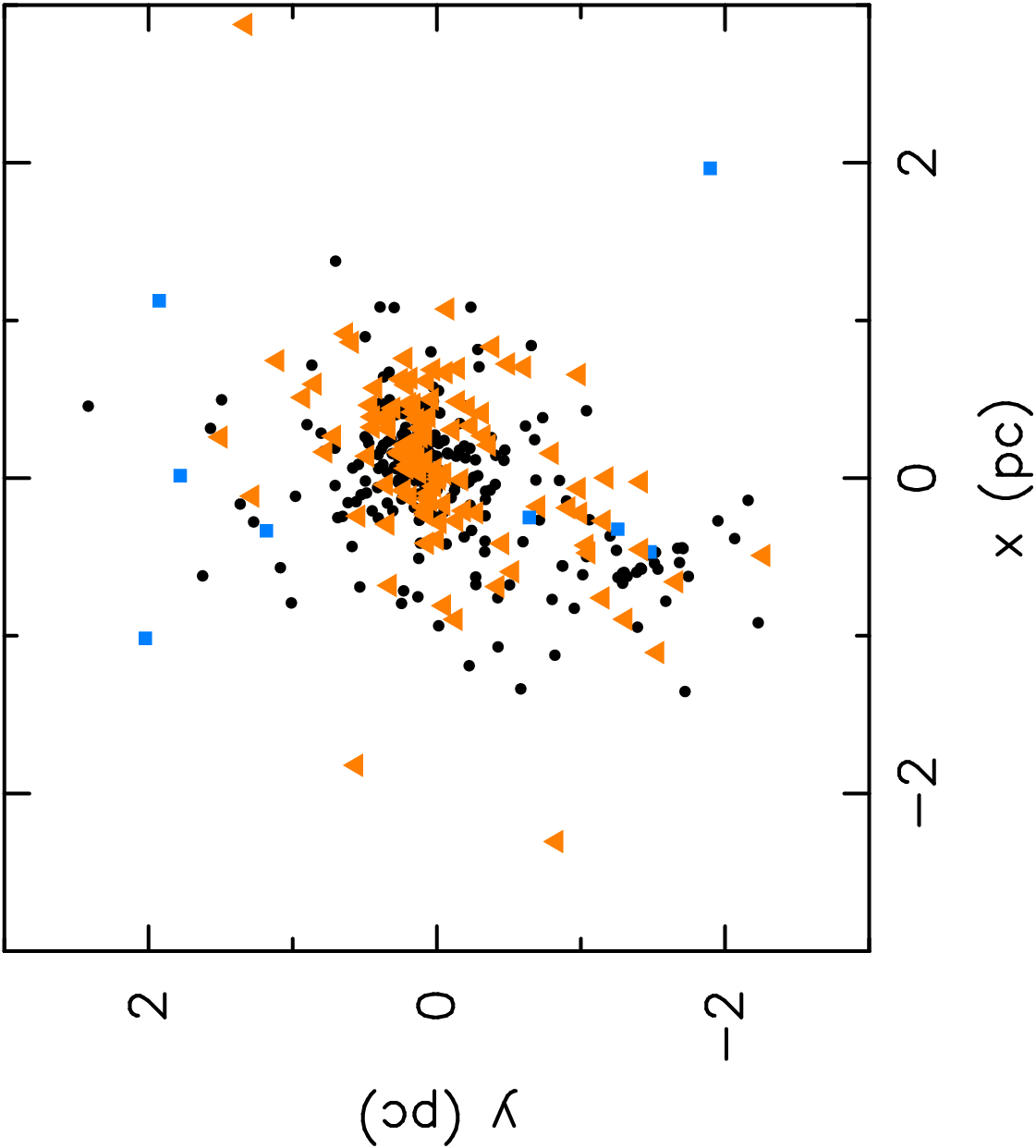}}} 
\hspace*{0.3cm}\subfigure[]{\label{Sim_05-b}\rotatebox{270}{\includegraphics[scale=0.35]{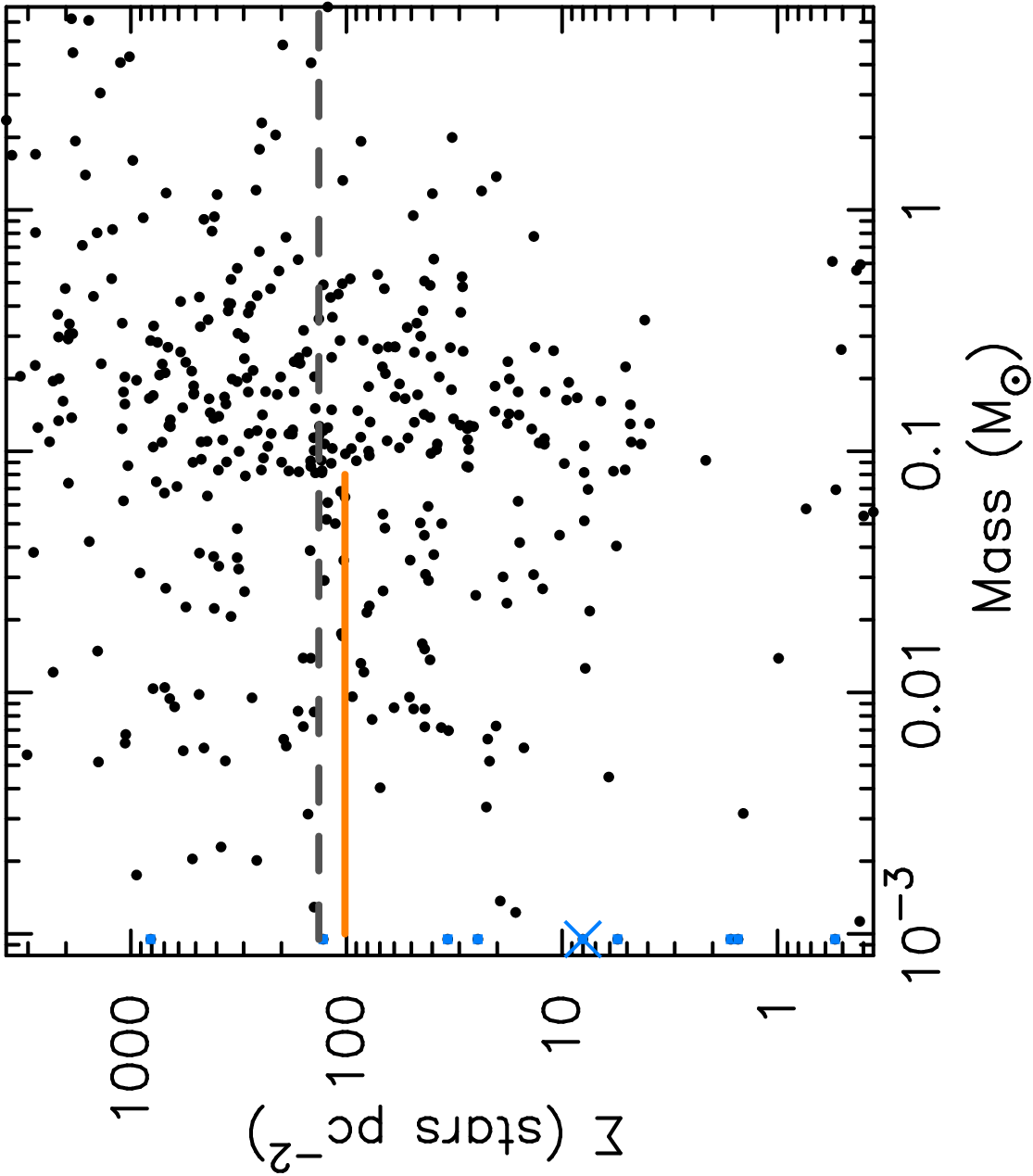}}}
\hspace*{-1.5cm} \subfigure[]{\label{Sim_05-c}\rotatebox{270}{\includegraphics[scale=0.35]{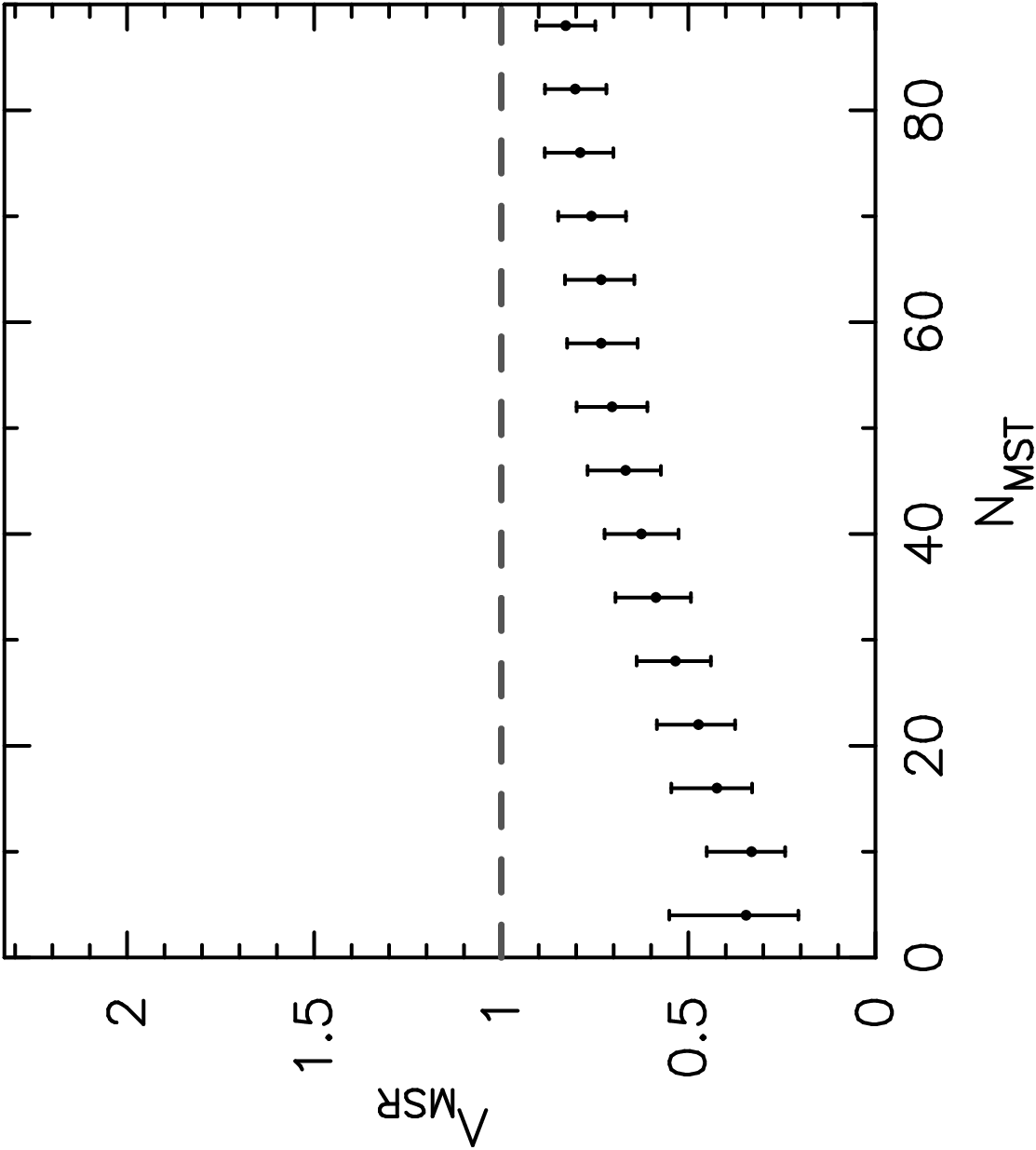}}} 
\hspace*{0.3cm}\subfigure[]{\label{Sim_05-d}\rotatebox{270}{\includegraphics[scale=0.35]{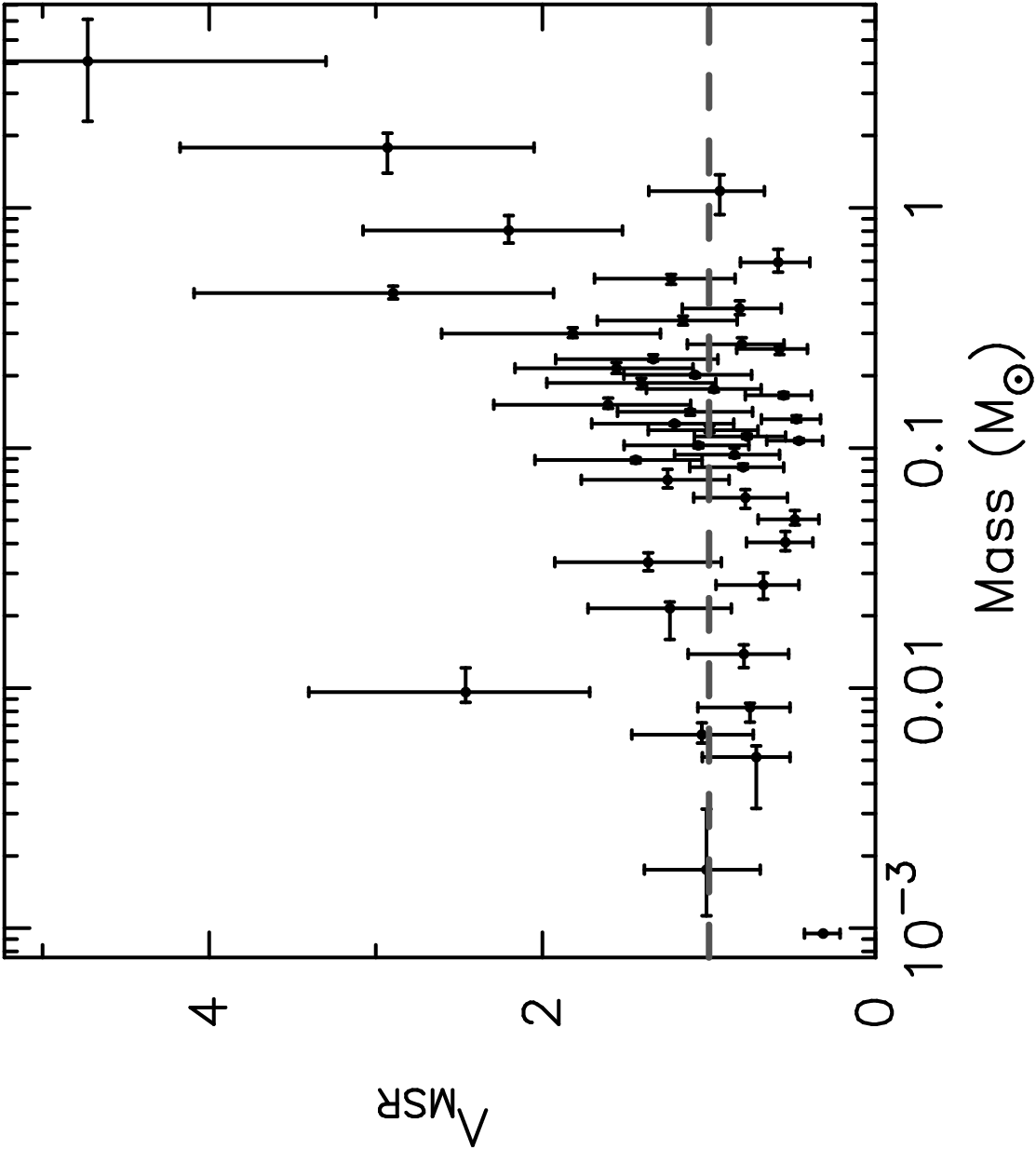}}}

\caption[bf]{A snapshot of the spatial distribution of stars, brown dwarfs and planetary-mass objects in one of our simulations in which the number of free-floating planets at 3\,Myr is similar to the number of H-type objects in IC\,348. The number of free-floating planets in this particular simulation is shown by the cyan line in panel (b) of the plot of the number of free-floating planets in Fig.~\ref{FFLOP_numbers}. In panel (a) we show a snapshot of the simulation, with stellar mass objects shown in black, brown dwarfs shown in orange, and free-floating planets shown in blue. In panel (b) we show the surface density of each object plotted against its mass; the median surface density of the free-floating planets is shown by the blue cross, the median surface density of brown dwarfs is shown by the orange horizontal line, and the median surface density of all objects is shown by the horizontal grey dashed line. In panel (c) we show the $\Lambda_{\rm MSR}$ mass segregation ratio for the lowest-mass objects, with each point being the determination of  $\Lambda_{\rm MSR}$ for the $N_{\rm MST}$ least massive objects. In panel (d) we show  $\Lambda_{\rm MSR}$ for discrete subsets of objects. In both panel (c) and panel (d) the horizontal dashed line indicates  $\Lambda_{\rm MSR} = 1$, which is the value when the objects are not mass segregated.    }
\label{Sim_05}
  \end{center}
\end{figure*}

\subsection{Spatial distribution of H-type objects in $N$-body simulations}

In the simulations that produce a similar number of free-floating planets to the number of H-type objects in IC\,348, we quantify the spatial distribution of the planets that remain within 5\,pc of the centre of the simulated star-forming regions.

In Fig.~\ref{Sim_05} we show a representative simulation in which the planets were all originally 5\,au from their host stars. This particular simulation is the cyan line in Fig.~\ref{FFLOP_numbers-b}, which has 9 free-floating planets after 3\,Myr (many more have already escaped the region).

In Fig.~\ref{Sim_05-a} we show an $x-y$ projection of the simulation after 3\,Myr. Brown dwarfs are shown by the orange triangles and free-floating planets are shown by the blue squares. In this snapshot the  free-floating planets appear more spread out, and this is confirmed by both the $\Sigma - m$ surface density--mass plot, and the $\Lambda_{\rm MSR}$ mass segregation ratio plots.

In Fig.~\ref{Sim_05-b} we show the $\Sigma - m$ plot; the median surface density for all objects is shown by the grey horizontal dashed line and the median surface density for all brown dwarfs (down to $1 \times 10^{-3}$\,M$_\odot$) is shown by the horizontal solid orange line. All of the free-floating planets have the same mass (1\,M$_{\rm Jup} = 9.4 \times 10^{-4}$\,M$_\odot$) and their median surface density is shown by the blue cross. The free-floating planets have significantly lower surface densities than both the stars, and the brown dwarfs in the region.

The surface density of all objects in this snapshot is 134\,stars\,pc$^{-2}$, whereas the free-floating planets have a surface density of 5.5\,stars\,pc$^{-2}$, and a KS-test returns $D = 0.54$ with a p-value of $6 \times 10^{-3}$ that they share the same underlying parent distribution. On contrast, the brown dwarfs have lower surface densities than the stars  (102\,stars\,pc$^{-2}$), but the KS test returns a p-value of 0.39, meaning we cannot reject the hypothesis that the brown dwarfs share the same surface density distribution as the stars. 

The $\Lambda_{\rm MSR}$ mass segregation ratio displays similar behaviour. In Fig.~\ref{Sim_05-c} we show the evolution of the $\Lambda_{\rm MSR}$ ratio for successively higher $N_{\rm MST}$ objects. We start with a bin size of $N_{\rm MST} = 4$ lowest mass objects, then add a further 6 objects to each sucessive bin. Clearly, the  Fig.~\ref{Sim_05-c} shows a significant inverse mass segregation signature in the lowest mass objects ($\Lambda_{\rm MSR} << 1$ i.e.\,\,they have a wider spatial distribution than the stars and brown dwarfs in the simulation). Similarly, when we calculate $\Lambda_{\rm MSR}$ for discrete mass bins (Fig.~\ref{Sim_05-d}) the very lowest mass objects have an inverse mass segregation signature (the leftmost bin). \\

Out of eight realisations of the same simulation initial conditions ($r_F = 1.3$\,pc, $D = 1.6$, $a_p = 5$\,au) that produce a similar number of free-floating planets to the number of H-type objects in IC\,348, the free-floating planets are significantly more dipsersed after 1\,Myr in all eight simulations.  After 3\,Myr the signal has disappeared in one simulation and is marginal in three others. After 5\,Myr the signal is only significant in four out of eight simulations.  

Therefore, whilst it is possible that the H-type objects in IC\,348 could be ejected free-floating planets, our simulation results strongly suggest that they would have a more dispersed spatial distribution, and lower surface densities, than both the stars and the brown dwarfs, which is not observed in IC\,348.

\section{Discussion}
\label{sec:discuss}

We have set up $N$-body simulations of the dynamical evolution of IC\,348 and have asked the question that if the H-type objects were originally planets that formed in a disc around stars, what initial conditions (in terms of stellar density, but also the initial semimajor axis distribution of the planets) would be required to produce a similar number of free-floating planets to the number of H-type objects (9) observed by \citet{Luhman25} in IC\,348.

In the simulations where we match the density and structure to the values observed in IC\,348, the majority of simulations with planets at 1\,au do not produce a significant (usually fewer than 4) number of free-floating planets, and the simulations in which the planets are originally at 30\,au produce too many free-floating planets (between 20 and 40). Simulations with the planets initially at 5\,au produce between 5 and 15 free-floating planets. We therefore posit that if the H-type objects were planets that formed around stars, they would have formed at a similar semimajor axis to Jupiter's current position in our Solar System.

However, when we quantify the spatial distribution of the free-floating planets in our simulations, we find that they typically are more dispersed than the stars and brown dwarfs in the simulations, and they also reside in areas of lower surface density than the stars and brown dwarfs. The only caveat to this is if the age of IC\,348 has been underestimated, as after around 5\,Myr the free-floating planets are either ejected from the region (and do not feature in the determination of the spatial distribution of free-floating planets), and/or the free-floating population has dynamically relaxed such that their spatial distributions are indistinguishable from those of the stars and brown dwarfs in half of our simulations.

We therefore conclude that the H-type objects in IC\,348 -- which have a spatial distribution that is indistinguishable from the stars and brown dwarfs -- likely formed in a similar way to stars and brown dwarfs. We note that there are many ideas for the formation of brown dwarfs, from the ejection of a stellar embryo from a core \citep{Reipurth01}, fragmentation of circumstellar discs \citep{Stamatellos07b,Stamatellos09}, photoerosion of a protostellar core by massive stars \citep{Whitworth04,Diamond24}, disruption of binary of triple systems \citep{Goodwin07b} to the idea that they simply form like stars from the collapse and fragmentation of giant molecular clouds \citep{Bate09}.

The photoerosion \citep{Whitworth04} and `star-like'  \citep{Bate09} formation mechanisms predict a similar spatial distribution to stars, so it is possible that the H-type objects may not form in exactly the same way as the more massive stars, but still have the same spatial distribution. By comparing our $N$-body simulations to the observations of IC\,348 we certaintly do not see any evidence that the H-type objects form like planets in bound orbits around stars and are then subsequently ejected by dynamical encounters.

In addition to the uncertainty around the age of the IC\,348 region described above, we highlight several further caveats.

First, whilst the best-fitting simulations to the median density ($\tilde{\Sigma}$) and structure ($\mathcal{Q}$-parameter) are consistent with the observations, the $\mathcal{Q}$ parameter values from all but one simulation are slightly higher than the value in IC\,348 (1.0 -- 1.2, compared to the observed value of 0.98). The $\mathcal{Q}$-parameter increases the more centrally concentrated a distribution is, meaning that if more distant objects are included the value becomes higher. This is an obvious problem when comparing simulations to observations, as the observations may be limited by certain incompleteness issues, whereas we know the location of every object in the simulations. In our analysis we limited the simulation data to the central 5\,pc, but reducing this radius will also reduce the  $\mathcal{Q}$-parameter. The determination of the median local density is more robust against changes to the field of view (both in simulations and observations).

The simulations do not include a primordial stellar binary population, nor do they include a background gas potential. The effects of gas removal on the dynamical evolution of star-forming regions are minimal \citep{Goodwin09,Kruijssen12a,Lucas20,Calovic25}, either due to high local star formation efficiencies \citep[which limits the amount of leftover gas, e.g.][]{Dale12b,Kruijssen12a}, gas being removed  through cavities in the region \citep{Lucas20} and subvirial stellar velocities, which results in a high \emph{effective} star-formation efficiency and increases the boundedness of the stars \citep{Goodwin09,Calovic25}.

The primordial stellar binary population in star-forming regions is very poorly constrained \citep{Rawcliffe25}, and may or may not be similar to the very much better determined binary population in the Galactic field \citep{Duquennoy91,Raghavan10,Ward-Duong15}. The presence of stellar binaries is likely to increase the number of interactions in the simulations, either via direct fly-bys or through secondary effects such as the von Zeipel-Lidov-Kozai mechanism \citep{vonZeipel10,Lidov62,Kozai62}. Therefore, the amount of free-floating planets produced in our simulations is likely an underestimate, as several authors have shown that binary systems are likely to be a rich production site for free-floating planets \citep{Malmberg07a,Coleman24}.

We assume that every star (0.1 -- 3\,M$_\odot$) hosts a single Jupiter-mass planet at the same semimajor axis (1, 5 or 30\,au, depending on the simulation). In reality, these planetary systems may contain more or fewer gas giants, and the systems with more than one gas giant that are subject to disruption may produce more than one free-floating planet \citep{Malmberg07a,FlamminiDotti19}. Additionally, the initial semimajor axis distribution is likely to be skewed towards smaller separations (and smaller planetary masses) based on the observed dependences of disc mass and disc radius on star mass \citep[e.g.][]{Coleman22}.

Finally, the simulations assume instantaneous star and planet formation. Whilst gas giant planets are thought to form very quickly \citep[within 1\,Myr,][]{Alves20,SeguraCox20} some of the dynamical encounters that eject planets happen on these timescales (as shown by the production of free-floating planets in Fig.~\ref{FFLOP_numbers}).

\section{Conclusions}
\label{sec:conclude}

We quantify the spatial distribution of the recently characterised spectral H-type substellar objects in IC\,348, and compare this with the spatial distribution of free-floating planets and brown dwarfs in $N$-body simulations tailored to match the present-day density and structure of IC\,348. Our conclusions are the following:

(i) The H-type objects do not have a different spatial distribution to the other substellar objects, or the stars, in IC\,348. They have a similar spatial distribution as quantified by the $\Lambda_{\rm MSR}$ mass segregation ratio, and they have a similar density distribution, as quantified by the $\Sigma - m$ local surface density--mass distribution. 

(ii) We constrain the likely initial conditions of IC\,348 by comparing snapshots of $N$-body simulations to the observed spatial distribution (the median local density $\tilde{\Sigma}$ and the morphology quantified by the $\mathcal{Q}$-parameter) to establish the likely initial semimajor axis distribution of planets that become free-floating in the simulations. A similar number of free-floating objects (5 -- 15) as the number of H-type objects in IC\,348 are produced when the initial semimajor axes of the planets are around 5\,au.

(iii) We then quantify the spatial distribution of the free-floating planets that remain within 5\,pc of the centre of the simulation at ages similar to that of IC\,348 (3\,Myr). The planets always have a much more dispersed spatial distribution (according to the $\Lambda_{\rm MSR}$ mass segregation ratio) and much lower surface densities (according to $\Sigma - m$) than the stars and brown dwarfs in the simulations.

(iv) We therefore suggest that the H-type objects represent a new type of brown dwarf, that formed via (a) mechanism(s) that imprints a spatial distribution that is indistinguishable from stars, rather than forming as planets that are then ejected from their host stars by close encounters.

\section*{Data availability statement}

 The observational census, including mass estimates for all objects, can be found at \href{https://doi.org/10.5281/zenodo.19696461}{https://doi.org/10.5281/zenodo.19696461}.  The data used to produce the plots in this paper will be shared on reasonable request to the corresponding author.

\section*{Acknowledgments}

We are grateful to the anonymous referee for a helpful report. We thank Kevin Luhman for providing the full observational census of IC\,348 from \citet{Luhman16} with the additons from \citet{Luhman25}. RJP acknowledges support from the Royal Society in the form of a Dorothy Hodgkin Fellowship.

\bibliographystyle{mnras}
\bibliography{general_ref}

\appendix

\section{$\Lambda_{\rm MSR}$ in IC\,348 with masses assigned to H-type objects}

  \begin{figure*}
  \begin{center}
\setlength{\subfigcapskip}{10pt}
\hspace*{-1.5cm} \subfigure[]{\label{IC348_Lambda_all_masses-a}\rotatebox{270}{\includegraphics[scale=0.35]{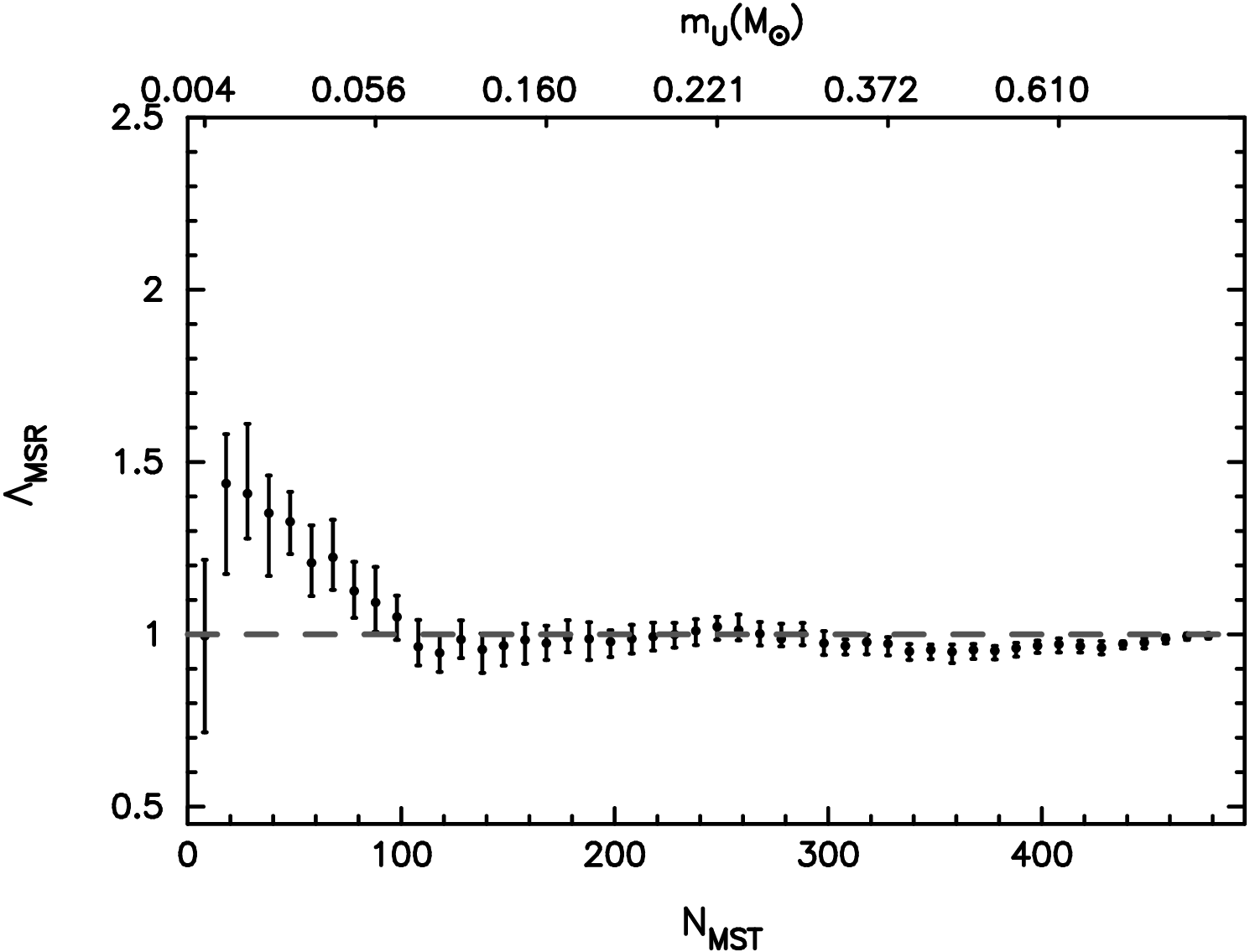}}} 
\hspace*{0.3cm}\subfigure[]{\label{IC348_Lambda_all_masses-b}\rotatebox{270}{\includegraphics[scale=0.35]{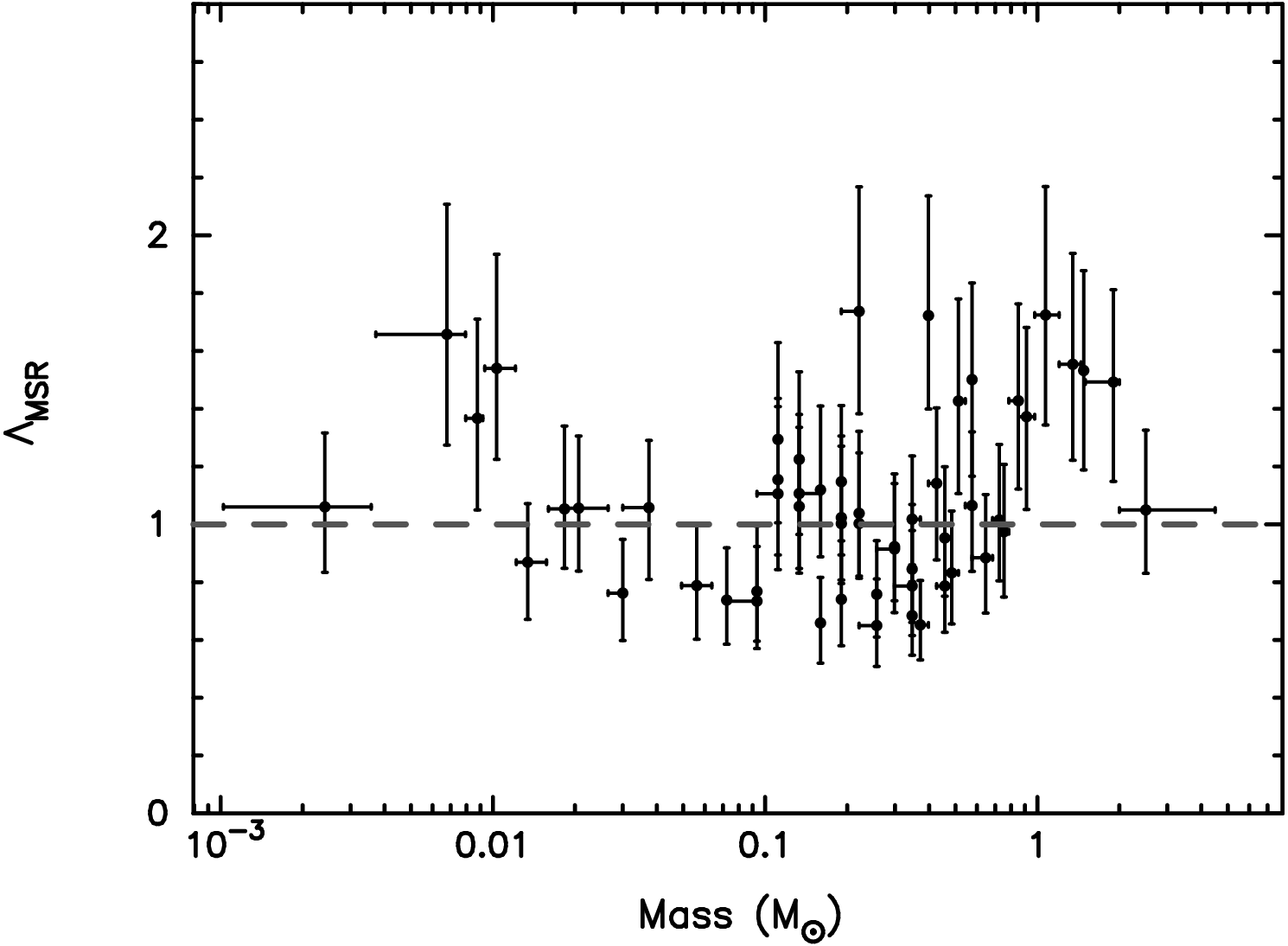}}}

\caption[bf]{As Fig.~\ref{IC348_Lambda} but where we retain the mass determination of the H-type objects from \citet{Luhman25}. In panel (a) we show the $\Lambda_{\rm MSR}$ ratio as a function of the $N_{\rm MST}$ least massive objects -- first constructing a bin containing the nine least-massive objects and then adding the next nine next-least massive objects. In panel (b) we show the   $\Lambda_{\rm MSR}$ ratio for fixed bins containing nine objects. Here, the error bars in the $x$-axis are the mass range of each bin. In both panels, the error bars in the $y$-axis are the uncertainties in the determination of $\Lambda_{\rm MSR}$, as described in the text. The horizontal dashed line indicates $\Lambda_{\rm MSR} = 1$, i.e. no mass segregation. }
\label{IC348_Lambda_all_masses}
  \end{center}
\end{figure*}

The $\Lambda_{\rm MSR}$ ratio can be applied to any subset of objects, based on their mass, spectral type, luminosity, magnitude, etc. However, the H-type objects span a mass range of $2.2 \times 10^{-3}$ -- $9.3 \times 10^{-2}$\,M$_\odot$, which overlaps with the brown dwarfs of other spectral types in the region.

In Fig.~\ref{IC348_Lambda} we assigned the H-type objects all masses of $1 \times 10^{-3}$\,M$_\odot$, and found no difference(s) between the H-type objects and the brown dwarfs and stars. In Fig.~\ref{IC348_Lambda_all_masses} we give the H-type objects their original masses to test if some of the H-type objects (for example the lowest mass ones) have a different spatial distribution to the others (e.g.\,\,the higher mass ones), or whether in combination with similar-mass brown dwarfs they have a different distribution to (e.g.) stars.

As with Fig.~\ref{IC348_Lambda}, we find no difference between the spatial distribution of substellar objects and stars, and our intepretation of  Fig.~\ref{IC348_Lambda} in the main paper is not compromised by setting all the H-type object masses to the same value.

\label{lastpage}

\end{document}